\documentclass[twoside,reqno]{article}%
\usepackage{epsfig,cite}
\usepackage{amssymb,amsmath}
\usepackage{times}
\usepackage{amsmath}
\usepackage{amsfonts}
\usepackage{amssymb}
\usepackage{graphicx}%

\begin{document}

 \setlength{\parindent}{15pt}
 \setlength{\textwidth}{11.2cm}
 \setlength{\textheight}{46\baselineskip}
 \setlength{\oddsidemargin}{0in}
 \setlength{\evensidemargin}{0in}
 \pagestyle{headings}
 \sloppy \raggedbottom
 \setcounter{page}{1}


 \newpage\setcounter{figure}{0}
 \setcounter{equation}{0}
 \setcounter{footnote}{0}
 \setcounter{table}{0}
 \setcounter{section}{0}


\noindent{\small USC-06/HEP-B1\hfill\hfill hep-th/0601091 }

\begin{center}
{\LARGE Lectures on Twistors}\footnote{Lectures delivered at the
\textquotedblleft2005 Summer School on String/M Theroy\textquotedblright\ in
Shanghai, China, and the International Symposium QTS4, \textquotedblleft
Quantum Theory and Symmetries IV\textquotedblright, Varna, Bulgaria.}

{\vskip0.5cm}

\textbf{Itzhak Bars}

\bigskip

\textit{Department of Physics and Astronomy \break University of Southern
California, Los Angeles, CA 90089-0484, USA }

\bigskip

\textbf{Abstract}
\end{center}

In these lectures I will discuss the following topics

\begin{itemize}
\item Twistors in 4 flat dimensions.

\begin{itemize}
\item Massless particles, constrained phase space ($x^{\mu},p^{\mu}$) versus twistors.

\item Physical states in twistor space.
\end{itemize}

\item Introduction to 2T-physics and derivation of 1T-physics holographs and twistors.

\begin{itemize}
\item Emergent spacetimes \& dynamics, holography, duality.

\item Sp(2,R) gauge symmetry, constraints, solutions and (d,2).

\item Global symmetry, quantization and the SO(d,2) singleton.

\item Twistors for particle dynamics in $d$ dimensions, particles with mass,
relativistic, non-relativistic, in curved spaces, with interactions.
\end{itemize}

\item Supersymmetric 2T-physics, gauge symmetries \& twistor gauge.

\begin{itemize}
\item Coupling $X,P,g$, gauge symmetries, global symmetries.

\item Covariant quantization, constrained generators \& representations of
G$_{\text{super}}$.

\item Twistor gauge: supertwistors dual to super phase space. Examples in d=4,6,10,11.

\end{itemize}

\item Supertwistors and some field theory spectra in d=4,6.

\begin{itemize}
\item Super Yang-Mills d=4, N=4; Supergravity d=4, N=8.

\item Self-dual tensor supermultiplet and conformal theory in d=6.
\end{itemize}

\item Twistor superstrings

\begin{itemize}
\item $d+2$ view of twistor superstring in $d=4$.

\item Worldsheet anomalies and quantization of twistor superstring.

\item Open problems.
\end{itemize}
\end{itemize}

\newpage

\section{Twistors in d=4 flat dimensions\label{twistd4}}

A massless spinless relativistic particle in $4$ space-time dimensions is
described by the action
\begin{equation}
S\left(  x,p\right)  =\int d\tau\left(  \partial_{\tau}x^{\mu}p_{\mu}-\frac
{1}{2}ep_{\mu}p_{\nu}\eta^{\mu\nu}\right)  . \label{cov}%
\end{equation}
It has a gauge symmetry under the transformations $\delta_{\varepsilon
}e=\partial_{\tau}\varepsilon\left(  \tau\right)  ,\delta_{\varepsilon}x^{\mu
}=\varepsilon\left(  \tau\right)  p^{\mu}$, $\delta_{\varepsilon}p_{\mu}=0.$
The generator of the gauge symmetry is $p^{2}/2,$and it vanishes as a
consequence of the equation of motion for the gauge field $\delta S/\delta
e=p^{2}/2=0.$This equation of motion is interpreted as demanding that the
solution space must be gauge invariant (since the generator must vanish).

In the covariant quantization of this system one defines the physical states
as those that satisfy the constraint $p^{2}|\phi\rangle=0,$ so that they are
gauge invariant. A complete set of physical states is found in momentum space
$|k\rangle$ on which the gauge generator is simultaneously diagonal with the
momentum operator $p_{\mu}|k\rangle=|k\rangle k_{\mu},$ and $p^{2}%
|k\rangle=|k\rangle k^{2}=0.$ The probability amplitude of a physical state in
position space $\langle x|\phi\rangle=\phi\left(  x\right)  $ satisfies the
condition $\langle x|p^{2}|\phi\rangle=0$ which gives the Klein-Gordon
equation $\partial^{2}\phi\left(  x\right)  =0$. The general solution is a
superposition of plane waves, which are the probability amplitudes of physical
states with definite momentum
\begin{gather}
\text{General solution: }\phi\left(  x\right)  =\int\frac{d^{4}k}{\left(
2\pi\right)  ^{4}}\delta\left(  k^{2}\right)  \left[  a\left(  k\right)
e^{ik\cdot x}+h.c.\right] \label{kgsln}\\
\text{ Plane wave with definite momentum }k^{\mu}\text{: }\phi_{k}\left(
x\right)  =\langle x|k\rangle\sim e^{ik\cdot x},\;k^{2}=0. \label{planewave}%
\end{gather}
A similar treatment for spinning particles leads to the spinning free field
equations, such as the Dirac equation, Maxwell equation, linearized Einstein
equation, etc.

\subsection{Twistors\label{sectwistors}}

The following shows several ways of solving the constraint $p^{2}=0$ or
$k^{2}=0$ that enter in these equations%

\begin{equation}
p^{2}=0:%
\genfrac{}{}{0pt}{}{p^{0}=\pm\sqrt{\vec{p}^{2}}}{p^{-}=p_{\perp}^{2}/2p^{+}}%
\;\text{or\ }%
\genfrac{}{}{0pt}{}{p_{\alpha\dot{\beta}}=\pm(\lambda\lambda^{{\Large \dagger
}})_{\alpha\dot{\beta}}=\frac{1}{\sqrt{2}}p^{\mu}\left(  \sigma_{\mu}\right)
_{\alpha\dot{\beta}}}{{\small 2\times2~}\text{Hermitian{\small , rank 1, up to
phase }}\lambda\rightarrow e^{i\phi}\lambda}
\label{pm}%
\end{equation}
In the second form, the matrix $p_{\alpha\dot{\beta}}$ is constructed from two
complex numbers $\lambda_{1},\lambda_{2}$ that form a doublet of SL$\left(
2,C\right)  =$SO$\left(  3,1\right)  $%
\begin{equation}
p={\small \pm}\left(
\genfrac{}{}{0pt}{}{\lambda_{1}}{\lambda_{2}}%
\right)  \left(  \lambda_{1}^{\ast}\;\lambda_{2}^{\ast}\right)  ={\small \pm
}\left(
\genfrac{}{}{0pt}{}{\lambda_{1}\lambda_{1}^{\ast}}{\lambda_{2}\lambda
_{1}^{\ast}}%
\genfrac{}{}{0pt}{}{\lambda_{1}\lambda_{2}^{\ast}}{\lambda_{2}\lambda
_{2}^{\ast}}%
\right)  =\frac{1}{\sqrt{2}}\left(
\genfrac{}{}{0pt}{}{p^{0}+p^{3}}{p_{1}+ip_{2}}%
\genfrac{}{}{0pt}{}{p_{1}-ip_{2}}{p^{0}-p^{3}}%
\right)
\end{equation}
This has automatically zero determinant $\det\left(  p\right)  =\left(
\lambda_{1}\lambda_{1}^{\ast}\right)  \left(  \lambda_{2}\lambda_{2}^{\ast
}\right)  -\left(  \lambda_{1}\lambda_{2}^{\ast}\right)  \left(  \lambda
_{2}\lambda_{1}^{\ast}\right)  =0=\left(  p^{0}+p^{3}\right)  \left(
p^{0}-p^{3}\right)  -\left(  p_{1}-ip_{2}\right)  \left(  p_{1}+ip_{2}\right)
=p_{0}^{2}-\vec{p}^{2},$ which imposes the desired solution $p^{\mu}p_{\mu}=0$
automatically. Note that the overall phase $e^{i\phi}$ of $\lambda_{\alpha}$
drops out, so the matrix $p_{\alpha\dot{\beta}}$ really has only 3 real
parameters, as it should.

The reader is reminded of a bit of group theory for$~$SL$(2,C)=$SO$\left(
3,1\right)  ~$
\begin{align*}
\text{spinors}\text{:~\ }  &  \left\{
\begin{array}
[c]{l}%
\lambda_{{\Large \alpha}}\;\left(  \frac{1}{2},0\right) \\
\bar{\lambda}_{\dot{\alpha}}\equiv\lambda_{\dot{\alpha}}^{\dagger}\;\left(
0,\frac{1}{2}\right)
\end{array}
\right.  ,\;\text{invariant tensors:~\ }\left\{
\begin{array}
[c]{c}%
\varepsilon^{\alpha\beta}\text{~or }\varepsilon^{\dot{\alpha}\dot{\beta}%
}=\left(
\genfrac{}{}{0pt}{}{0}{-1}%
\genfrac{}{}{0pt}{}{1}{0}%
\right) \\
\text{{\small metric,~raise / lower indices}}%
\end{array}
\right. \\
\text{vectors}\text{:\ }  &  \left\{
\begin{array}
[c]{l}%
\left(  \sigma_{\mu}\right)  _{\alpha\dot{\beta}}=\left(  1,\vec{\sigma
}\right)  _{\alpha\dot{\beta}};\;\;\left(  \bar{\sigma}_{\mu}\right)
_{\dot{\alpha}\beta}=\left(  -1,\vec{\sigma}\right)  _{\dot{\alpha}\beta
};\;\;{\small (}\frac{{\small 1}}{{\small 2}}{\small ,}\frac{{\small 1}%
}{{\small 2}}{\small )}=%
\genfrac{}{}{0pt}{}{{\small (}\frac{{\small 1}}{{\small 2}}{\small ,0)\times
(0,\frac{1}{2})}}{{\small (0,\frac{1}{2})\times(}\frac{{\small 1}}{{\small 2}%
}{\small ,0}}%
~\\
p^{\mu}:~p_{\alpha\dot{\beta}}=\frac{1}{\sqrt{2}}p^{\mu}\left(  \sigma_{\mu
}\right)  _{\alpha\dot{\beta}},\;\bar{p}_{\dot{\alpha}\beta}=\frac{1}{\sqrt
{2}}p^{\mu}\left(  \bar{\sigma}_{\mu}\right)  _{\dot{\alpha}\beta},\;\\
x^{\mu}:\;x^{\alpha\dot{\beta}}=\frac{1}{\sqrt{2}}x^{\mu}\left(  \sigma_{\mu
}\right)  ^{\alpha\dot{\beta}},\;\bar{x}^{\dot{\alpha}\beta}=\frac{1}{\sqrt
{2}}x^{\mu}\left(  \bar{\sigma}_{\mu}\right)  ^{\dot{\alpha}\beta}%
\end{array}
\right.
\end{align*}
Penrose \cite{penrose}\cite{penrose2} suggested a second spinor $\mu
^{\dot{\alpha}}$ and introduced the \textquotedblleft incidence
relation\textquotedblright\ which defines $x$ as being roughly the
\textquotedblleft slope\textquotedblright\ of a \textquotedblleft
line\textquotedblright\ in spinor space%
\begin{equation}
\mu^{\dot{\alpha}}=-i~\bar{x}^{\dot{\alpha}\beta}\lambda_{\beta},\text{ a
\textquotedblleft line\textquotedblright\ in spinor space.} \label{line}%
\end{equation}
Finally a twistor is defined as $Z_{A}=\left(
\genfrac{}{}{0pt}{}{\mu^{\dot{\alpha}}}{\lambda_{\alpha}}%
\right)  ,\;A=1,2,3,4,$ that bundles together $\mu$ and $\lambda$ as a
quartet. If $\mu$ satisfies the Penrose relation, then the pair $\mu,\lambda$
is equivalent to the the phase space of the massless particle%
\begin{equation}
Z_{A}=\left(
\genfrac{}{}{0pt}{}{\mu^{\dot{\alpha}}}{\lambda_{\alpha}}%
\right)  =\left(
\genfrac{}{}{0pt}{}{\left(  -i\bar{x}\lambda\right)  ^{\dot{\alpha}}%
}{\lambda_{\alpha}}%
\right)  \Leftrightarrow%
\genfrac{}{}{0pt}{}{\text{on-shell}}{\text{ }\genfrac{}{}{0pt}{}{\text{phase}%
}{\text{space}}\text{~}\left(  x^{\mu},p_{\mu}\right)  }
\label{quartet}%
\end{equation}

Although not manifest, the massless particle action above has a hidden
conformal symmetry SO$\left(  4,2\right)  .$ This symmetry can be made
manifest through the twistor since SO$\left(  4,2\right)  =$ SU$\left(
2,2\right)  $ and the quartet $Z_{A}$ can be classified as the fundamental
representation $4$ of SU$\left(  2,2\right)  .$ This non-compact group has a
metric which can be taken as $C=\left(
\genfrac{}{}{0pt}{}{0}{1}%
\genfrac{}{}{0pt}{}{1}{0}%
\right)  =\sigma_{1}\times1.$ Using the metric we define the other fundamental
representation $\bar{4}$ of SU$\left(  2,2\right)  $ and relate it to the
complex conjugate of $Z_{A}$ as follows
\begin{equation}
\bar{Z}^{A}=Z^{\dagger}C=\left(  \lambda_{\dot{\alpha}}^{\dagger}%
\;~\mu^{\dagger\alpha}\right)  =\left(  \lambda_{\dot{\alpha}}^{\dagger
}~\;\left(  i\lambda^{\dagger}\bar{x}\right)  ^{\alpha}\right)  ,\;C=\sigma
_{1}\times1
\end{equation}
So $\bar{Z}^{A}Z_{A}$ is invariant under SU$\left(  2,2\right)  .$ We remind
the reader that the $4$ and \={4} of SU$\left(  2,2\right)  $ correspond to
the two Weyl spinors of SO$\left(  4,2\right)  .$ Now, with $\mu$ as given
above, we have
\begin{equation}
\bar{Z}^{A}Z_{A}=\lambda_{\dot{\alpha}}^{\dagger}\mu^{\dot{\alpha}}%
+\mu^{\dagger\alpha}\lambda_{\alpha}=-i\lambda^{\dagger}\bar{x}\lambda
+i\lambda^{\dagger}\bar{x}\lambda=0.\;
\end{equation}

So, by construction the $Z_{A}$ are 4 constrained complex numbers. But we can
reverse this reasoning, and realize that the definition of twistors is just
the statement that $Z_{A}$ is a quartet that has an overall irrelevant phase
and that is constrained by $\bar{Z}^{A}Z_{A}=0.$ Then the form of $\mu$ in
terms of $\lambda$ can be understood as one of the possible ways of
parameterizing a solution. The solution $\mu^{\dot{\alpha}}=-i~\bar{x}%
^{\dot{\alpha}\beta}\lambda_{\beta}$ is interpreted as the massless
particle. This is the conventional interpretation of twistors.

However, recently it has been realized that there are many other
ways of parameterizing solutions for the same $Z_{A}$ in terms of
phase spaces that have many other different interpretations
\cite{twistorBP1}. For any solution, if we count the number of
independent real degrees of freedom, we find%

\begin{equation}
\text{{\small Independent}}{\small :~}(8~\text{{\small real}}{\small ~}%
Z)-\left(
\genfrac{}{}{0pt}{}{\text{1 {\small real constraint}}}{\text{1 {\small real
phase}}}%
\right)  =6\text{ real}=%
\genfrac{}{}{0pt}{}{\text{{\small same as}}}{\text{3}\vec{x}\text{+3}\vec{p}}%
\end{equation}
This is the right number not only for the massless particle, but also the
massive particle, relativistic or non-relativistic, in flat space or curved
space, interacting or non-interacting.

Next, we compute the canonical structure for the pair $\left(  Z_{A},\bar
{Z}^{A}\right)  ,$ and we find that it is equivalent to the canonical
structure in phase space for the massless particle, iff we use the solution
$\mu^{\dot{\alpha}}=-i~\bar{x}^{\dot{\alpha}\beta}\lambda_{\beta}$
\begin{align*}
L &  =i\bar{Z}^{A}\partial_{\tau}Z_{A}=i\bar{\lambda}_{\dot{\alpha}}%
\partial_{\tau}\mu^{\dot{\alpha}}+i\bar{\mu}^{\alpha}\partial_{\tau}%
\lambda_{\alpha}\\
&  =\lambda_{\dot{\alpha}}^{\dagger}\partial_{\tau}\left(  \bar{x}%
\lambda\right)  ^{\dot{\alpha}}-\left(  \lambda^{\dagger}\bar{x}\right)
^{\alpha}\partial_{\tau}\lambda_{\alpha}\\
&  =\lambda_{\dot{\alpha}}^{\dagger}\left(  \partial_{\tau}\bar{x}%
^{\dot{\alpha}\beta}\right)  \lambda_{\beta}=Tr\left(  p\partial_{\tau}\bar
{x}\right)  =p_{\mu}\partial_{\tau}x^{\mu}%
\end{align*}
So the canonical pairs $\left(  Z_{A},i\bar{Z}^{A}\right)  $ or
$\left( \lambda_{\alpha},i\mu^{\dagger\alpha}\right)  $ or $\left(
x^{\mu},p_{\mu }\right)  $ are equivalent as long as they satisfy
the respective constraints $\bar{Z}^{A}Z_{A}=0$ and $p^{2}=0$. If we
use some of the other solutions given in \cite{twistorBP1} then the
correct canonical structure emerges for the massive particle, etc.,
all from the same twistor (see below).

Just like the constraint $p^{2}=0$ followed from an action principle in
Eq.(\ref{cov}), the constraints $\bar{Z}^{A}Z_{A}=0$ can also be obtained from
the following action principle by minimizing with respect to $V$
\begin{equation}
S\left(  Z\right)  =\int d\tau~\left(  \bar{Z}^{A}iD_{\tau}Z_{A}-2hV\right)
=\int d\tau~\left(  \bar{Z}^{A}i\partial_{\tau}Z_{A}+V\bar{Z}^{A}%
Z_{A}-2hV\right)  . \label{action}%
\end{equation}
Here $D_{\tau}Z_{A}=\partial_{\tau}Z_{A}-iVZ_{A}$ is a covariant derivative
for a U(1) gauge symmetry $Z_{A}\left(  \tau\right)  \rightarrow Z_{A}%
^{\prime}\left(  \tau\right)  =e^{i\omega\left(  \tau\right)
}Z_{A}\left( \tau\right)  .$ The gauge symmetry is precisely what is
needed to remove the unphysical overall phase noted above.

For spinning particles, an extra term $-2hV$ is included in the
action (missing in former literature). This term is gauge invariant
by itself under the U$\left(  1\right)  $ gauge transformation of
$V.$ We have been discussing the spinless particle $h=0,$ but
twistors can be generalized to spinning particles by taking
$h\neq0$. The equation of motion with respect to $V$ gives the
constraint $\bar{Z}^{A}Z_{A}=2h.$ If the twistor transform for
massless particles, appropriately modified to include
spin, is used to solve this constraint \cite{penrose}\cite{penrose2}%
\cite{shirafuji}, then $h$ is interpreted as the helicity of the spinning
massless particle. But if the more general transforms in \cite{twistorBP1} are
used, then $h$ is not helicity, but is an eigenvalue of Casimir operators of
SU$\left(  2,2\right)  $ in a representation for spinning
particles\footnote{This point will be discussed in detail in a future
paper.\label{future}}.

We have argued that the twistor action $S\left(  Z\right)  $ is
equivalent to the spinless massless particle action $S\left(
x,p\right)  $ (at least in one of the possible ways of
parameterizing its solutions). But note that $S\left( Z\right)  $ is
manifestly invariant under the global symmetry SU$(2,2)$. This is
the hidden conformal symmetry SO$(4,2)$ of the massless particle
action $S\left(  x,p\right)  $. Applying Noether's theorem we derive
the conserved current, which in turn is written in terms of
$x^{\mu},p_{\mu}$ as follows
\begin{align}
J_{A}^{~B}  &  =Z_{A}\bar{Z}^{B}-\frac{1}{4}Z_{C}\bar{Z}^{C}\delta_{A}%
^{~B}=\left(
\genfrac{}{}{0pt}{}{-i\bar{x}\lambda}{\lambda}%
\right)  \left(  \lambda^{\dagger}\;~i\lambda^{\dagger}\bar{x}\right) \\
&  =\left(
\genfrac{}{}{0pt}{}{-i\bar{x}\lambda\lambda^{\dagger}}{\lambda_{\alpha}%
\lambda_{\dot{\beta}}^{\dagger}}%
\genfrac{}{}{0pt}{}{\bar{x}\lambda\lambda^{\dagger}\bar{x}}{i\lambda_{\alpha
}\lambda^{\dagger}\bar{x}}%
\right)  =\left(
\genfrac{}{}{0pt}{}{i\bar{x}p}{p}%
\genfrac{}{}{0pt}{}{\bar{x}p\bar{x}}{-ip\bar{x}}%
\right)  =\frac{1}{4i}\Gamma^{MN}L_{MN}\label{ZZL}\\
&  =\frac{1}{2i}\left(  -\Gamma^{+^{\prime}-^{\prime}}L^{+^{\prime}-^{\prime}%
}+~\frac{1}{2}L_{\mu\nu}\Gamma^{\mu\nu}-\Gamma_{~\mu}^{+^{\prime}}%
L^{-^{\prime}\mu}-\Gamma_{~\mu}^{-^{\prime}}L^{+^{\prime}\mu}\right)
\end{align}
In the last line the traceless $4\times4$ matrix $\left(
\genfrac{}{}{0pt}{}{i\bar{x}p}{p}%
\genfrac{}{}{0pt}{}{\bar{x}p\bar{x}}{-ip\bar{x}}%
\right)  $ is expanded in terms of the following complete set of SO$\left(
4,2\right)  $ gamma matrices $\Gamma^{MN}\;$($M=\pm,\mu,$ see footnote
(\ref{gamms}))%
\begin{align}
\Gamma^{+^{\prime}-^{\prime}}  &  =\left(
\genfrac{}{}{0pt}{}{-1}{0}%
\genfrac{}{}{0pt}{}{0}{1}%
\right)  ,\ \Gamma^{\mu\nu}=\left(
\genfrac{}{}{0pt}{}{\bar{\sigma}^{\mu\nu}}{0}%
\genfrac{}{}{0pt}{}{0}{\sigma^{\mu\nu}}%
\right)  ,\;%
\genfrac{}{}{0pt}{}{\bar{\sigma}^{\mu\nu}\equiv\bar{\sigma}^{[\mu}\sigma
^{\nu]}}{\sigma^{\mu\nu}\equiv\sigma^{\lbrack\mu}\bar{\sigma}^{\nu]}}%
\label{gam1}\\
\Gamma^{+^{\prime}\mu}  &  =i\sqrt{2}\left(
\genfrac{}{}{0pt}{}{0}{0}%
\genfrac{}{}{0pt}{}{\bar{\sigma}^{\mu}}{0}%
\right)  ,\ \Gamma^{-^{\prime}\mu}=-i\sqrt{2}\left(
\genfrac{}{}{0pt}{}{0}{\sigma^{\mu}}%
\genfrac{}{}{0pt}{}{0}{0}%
\right)  ,\ \label{gam2}%
\end{align}
This identifies the generators of the conformal group $L^{MN}$ as the
coefficients%
\begin{equation}
L^{+^{\prime}-^{\prime}}=x\cdot p,\;L^{\mu\nu}=x^{\mu}p^{\nu}-x^{\nu}p^{\mu
},\;L^{+^{\prime}\mu}=p^{\mu},\;L^{-^{\prime}\mu}=\frac{x^{2}}{2}p^{\mu
}-x^{\mu}x\cdot p.
\end{equation}
It can be checked that this form of $L^{MN}$ are the generators of the hidden
SO$\left(  4,2\right)  $ conformal symmetry of the massless particle action.
The SO$\left(  4,2\right)  $ transformations are given by the Poisson brackets
$\delta x^{\mu}=\frac{1}{2}\omega_{MN}\left\{  L^{MN},x^{\mu}\right\}  $ and
$\delta p^{\mu}=\frac{1}{2}\omega_{MN}\left\{  L^{MN},p^{\mu}\right\}  ,$ and
these $L^{MN}$ are the conserved charges given by Noether's theorem.
Furthermore they obey the SO$\left(  4,2\right)  $ Lie algebra under the
Poisson brackets. This result is not surprising once we have explained that
$S\left(  Z\right)  =S\left(  x,p\right)  $ via the twistor transform.

The same SU$\left(  2,2\right)  $ symmetry of the twistor action $S\left(
Z\right)  $ has other interpretations as the hidden symmetry of an assortment
of other particle actions when other forms of twistor transform is used, as
explained in \cite{twistorBP1}. This recent broader result may seem surprising
because it is commonly unfamiliar.

\subsection{Physical states in twistor space}

In covariant quantization a physical state for a particle of any
helicity should satisfy the helicity constraint
$\frac{1}{2}(Z_{A}\bar{Z}^{A}+\bar
{Z}^{A}Z_{A})|\psi\rangle=2h|\psi\rangle.$ This is interpreted as
meaning that the physical state $|\psi\rangle$ is invariant under
the U$\left(  1\right)  $ gauge transformation generated by the
constraint that followed from the twistor action $S\left(  Z\right)
$. The probability amplitude in $Z$ space is
$\psi\left(  Z\right)  \equiv\langle Z|\psi\rangle$, so we can write $\bar{Z}%
^{A}\psi\left(  Z\right)  =\langle Z|\bar{Z}^{A}|\psi\rangle=-\frac
{\partial}{\partial Z_{A}}\psi\left(  Z\right).$ Then the helicity
constraint $\frac{1}{2}\langle Z|(Z_{A}\bar{Z}^{A}+\bar{Z}^{A}Z_{A}%
)|\psi\rangle=2h\langle Z|\psi\rangle$ produces the physical state condition,
\begin{equation}
Z_{A}\frac{\partial}{\partial Z_{A}}\psi\left(  Z\right)  =\left(
-2h-2\right)  \psi\left(  Z\right)  \label{htwistor}%
\end{equation}
for a particle of helicity $h.$ So a physical wavefunction in twistor space
$\psi\left(  \lambda,\mu\right)  $ that describes a particle with helicity $h$
must be homogeneous of degree $\left(  -2h-2\right)  $ under the rescaling
$Z\rightarrow tZ$ or $\left(  \mu,\lambda\right)  \rightarrow\left(
t\mu,t\lambda\right)  $ \cite{penrose}\cite{penrose2}. This is the only
requirement for a physical state $\psi\left(  Z\right)  $ in twistor space,
and it is easily satisfied by an infinite set of functions.

If we use the twistor transform for massless particles $\mu=-i\bar{x}\lambda$
and $p=\lambda\bar{\lambda},$ then any homogeneous physical state in twistor
space should be a superposition of massless particle wavefunctions since
$p^{2}=0$ is automatically satisfied. A similar statement would hold for any
of the other twistor transforms given in \cite{twistorBP1}, so a physical
state in twistor space can also be expanded in terms of the
wavefunctions$^{{\small \ref{future}}}$ of the particle systems discussed in
\cite{twistorBP1}\cite{twistorBP2}.

Let us now consider the expansion of a physical state $|\psi\rangle$
in terms of momentum eigenstates
$p^{\mu}|k\rangle=k^{\mu}|k\rangle,$ for a massless particle with
$k^{2}=0.$ We parameterize $k_{\alpha\dot{\beta}}=\pi_{\alpha
}\bar{\pi}_{\dot{\beta}}$ as in Eq.(\ref{pm}), where $\pi_{\alpha} $
can be redefined up to a phase $\pi\rightarrow e^{i\gamma}\pi$
without changing the physical state $|k\rangle$. In position space
such a physical state gave the plane wave as in
Eq.(\ref{planewave}), which we can rewrite as $\phi _{k}\left(
x\right)  =\langle x|k\rangle\sim\exp\left(  ik\cdot x\right)
=\exp\left(  iTr\bar{x}\pi\bar{\pi}\right)  .$ The twistor space
analog is $\phi_{k}\left(  \lambda,\mu\right)  =\langle
Z|k\rangle=\langle\lambda ,\mu|\pi,\bar{\pi}\rangle$. Since
$|k\rangle$ is a complete set of states, it is possible to write a
general physical state in twistor space as an infinite superposition
of the $\langle Z|k\rangle$ with arbitrary coefficients, in the same
way as the general solution of the Klein-Gordon equation in
Eq.(\ref{kgsln})%
\begin{equation}
\psi\left(  Z\right)  =\int d^{2}\pi d^{2}\bar{\pi}\left[  a\left(  \pi
,\bar{\pi}\right)  \langle Z|\pi,\bar{\pi}\rangle+h.c.\right]  \label{psipi}%
\end{equation}

To determine $\langle Z|k\rangle=\langle\lambda,\mu|\pi,\bar{\pi}\rangle$,
first note that the eigenstate of $\lambda_{\alpha}$ is proportional to
$\pi_{\alpha}$, so there must be an overall delta function $\langle\lambda
,\mu|\pi,\bar{\pi}\rangle\sim\delta\left(  \langle\lambda\pi\rangle\right)  .$
The argument of the delta function is the SL$\left(  2,C\right)  $ invariant
dot product defined by the symbol $\langle\lambda\pi\rangle\equiv
\lambda_{\alpha}\pi_{\beta}\varepsilon^{\alpha\beta}.$ The vanishing of
$\langle\lambda\pi\rangle=0$ requires $\lambda_{\alpha}\propto\pi_{\alpha},$
hence in the wavefunction $\langle\lambda,\mu|\pi,\bar{\pi}\rangle$ we can
replace $\lambda_{\alpha}=\frac{\lambda}{\pi}\pi_{\alpha}$ up to an overall
constant $c$ symbolized by $c=\frac{\lambda}{\pi}.$ This is the ratio of
either component $\frac{\lambda}{\pi}\equiv\frac{\lambda_{1}}{\pi_{1}}%
=\frac{\lambda_{2}}{\pi_{2}}.$ So we can write $\langle Z|k\rangle
=\langle\lambda,\mu|\pi,\bar{\pi}\rangle=$ $\delta\left(  \langle\lambda
\pi\rangle\right)  f\left(  \pi,\bar{\pi},\frac{\lambda}{\pi},\mu\right)  .$
Next examine the matrix elements of the twistor transform $p_{\alpha\dot
{\beta}}-\lambda_{\alpha}\bar{\lambda}_{\dot{\beta}}=0$ and apply the
operators on either the ket or the bra as follows ($\bar{\lambda}_{\dot{\beta
}}$ acts as a derivative $-\frac{\partial}{\partial\mu^{\dot{\beta}}}$ on the
eigenvalue of $\mu^{\dot{\beta}}$)
\begin{align}
0  &  =\langle Z|\left(  p_{\alpha\dot{\beta}}-\lambda_{\alpha}\bar{\lambda
}_{\dot{\beta}}\right)  |k\rangle=\left(  k_{\alpha\dot{\beta}}+\lambda
_{\alpha}\frac{\partial}{\partial\mu^{\dot{\beta}}}\right)  \langle\lambda
,\mu|\pi,\bar{\pi}\rangle\\
&  =\delta\left(  \langle\lambda\pi\rangle\right)  \pi_{\alpha}\left(
\bar{\pi}_{\dot{\beta}}+\frac{\lambda}{\pi}\frac{\partial}{\partial\mu
^{\dot{\beta}}}\right)  f\left(  \pi,\bar{\pi},\frac{\lambda}{\pi},\mu\right)
.
\end{align}
The solution is $f\left(  \pi,\bar{\pi},\lambda,\mu\right)  =g\left(  \pi
,\bar{\pi},\frac{\lambda}{\pi}\right)  \exp\left(  -\frac{\pi}{\lambda}%
\bar{\pi}_{\dot{\alpha}}\mu^{\dot{\alpha}}\right)  $, for any $g\left(
\pi,\bar{\pi},\frac{\lambda}{\pi}\right)  ,$ so $\langle Z|k\rangle
=\delta\left(  \langle\lambda\pi\rangle\right)  \exp\left( - \frac{\pi}%
{\lambda}\bar{\pi}_{\dot{\alpha}}\mu^{\dot{\alpha}}\right)  g\left(  \pi
,\bar{\pi},\frac{\lambda}{\pi}\right)  .$ Note that the exponential is a
rewriting of the plane wave $\exp\left(  iTr\bar{x}\pi\bar{\pi}\right)  $ by
using $\pi=\frac{\pi}{\lambda}\lambda$ and then setting $\mu=-i\bar{x}%
\lambda.$

Finally we determine $g\left(  \pi,\bar{\pi},\frac{\lambda}{\pi}\right)  $ for
a particle with any helicity $h$. According to the previous paragraph, since
$\langle Z|k\rangle$ is a physical wavefunction, it should be homogeneous of
degree $\left(  -2h-2\right)  $ under a rescaling $\left(  \mu,\lambda\right)
\rightarrow\left(  t\mu,t\lambda\right)  .$ It should also be phase invariant
under the phase transformations $\pi\rightarrow e^{i\gamma}\pi,$ $\bar{\pi
}\rightarrow e^{-i\gamma}\bar{\pi}$ since the momentum state $|k\rangle$
labeled by $k_{\alpha\dot{\beta}}=\pi_{\alpha}\bar{\pi}_{\dot{\beta}}$ is
phase invariant. The exponential $\exp\left(  -\frac{\pi}{\lambda}\bar{\pi
}_{\dot{\alpha}}\mu^{\dot{\alpha}}\right)  $ is homogeneous as well as phase
invariant, while the delta function satisfies $\delta\left(  \langle t\lambda
e^{i\gamma}\pi\rangle\right)  =t^{-1}e^{-i\gamma}\delta\left(  \langle
\lambda\pi\rangle\right)  .$ These considerations determine $g\left(  \pi
,\bar{\pi},\frac{\lambda}{\pi}\right)  =\left(  \frac{\lambda}{\pi}\right)
^{-1-2h}\phi_{h}\left(  \pi,\bar{\pi}\right)  ,$ with $\phi_{h}\left(
e^{i\gamma}\pi,e^{-i\gamma}\bar{\pi}\right)  =e^{-i2h\gamma}\phi_{h}\left(
\pi,\bar{\pi}\right)  .$

The specific $\phi_{h}\left(  \pi,\bar{\pi}\right)  $ for each helicity are
determined as follows. $\phi_{h}\left(  \pi,\bar{\pi}\right)  $ must have
SL$\left(  2,C\right)  $ spinor indices for the representation $\left(
j_{1},j_{2}\right)  ,$since for a spinning particle the complete set of labels
includes Lorentz indices $|k,j_{1},j_{2},\cdots\rangle$ in addition to
momentum$.$ The chirality of the SL$\left(  2,C\right)  $ labels must be
compatible with the spin $j_{1}+j_{2}=\left\vert h\right\vert $. So this
determines the Lorentz indices on the wavefunction $\phi_{h}\left(  \pi
,\bar{\pi}\right)  $ as well as the coefficients $a\left(  \pi,\bar{\pi
}\right)  $ in Eq.(\ref{psipi}). Examples of the overall wavefunction $\langle
Z|k\rangle$ is given in the table below%
\begin{equation}%
\begin{tabular}
[c]{||l|l|l||}\hline\hline
\multicolumn{3}{||l||}{$\;\;\langle Z|k\rangle=\delta\left(  \langle\lambda
\pi\rangle\right)  \exp\left( - \frac{\pi}{\lambda}\bar{\pi}_{\dot{\alpha}}%
\mu^{\dot{\alpha}}\right)  \left(  \frac{\lambda}{\pi}\right)  ^{-1-2h}%
\phi_{h}\left(  \pi,\bar{\pi}\right)  .$}\\\hline\hline
particle & $~\left(  j_{1},j_{2}\right)  $ & $~\;\phi_{h}\left(  \pi,\bar{\pi
}\right)  $\\\hline\hline
scalar & $~\left(  0,0\right)  $ & $~\;\phi_{0}\left(  \pi,\bar{\pi}\right)
=1$\\\hline
$\text{quark}$ & $%
\begin{array}
[c]{c}%
(0,\frac{1}{2})\\
(\frac{1}{2},0)
\end{array}
$ & $%
\begin{array}
[c]{c}%
\psi_{\dot{\alpha}}^{h=+1/2}\left(  \pi,\bar{\pi}\right)  =\bar{\pi}%
_{\acute{\alpha}}\\
\psi_{\alpha}^{h=-1/2}\left(  \pi,\bar{\pi}\right)  =\pi_{\alpha}%
\end{array}
$\\\hline
$%
\genfrac{}{}{0pt}{}{\text{gauge potential}}{A_{\mu}\;\;\;\;\;\;\;\;\;\;\;\;\;}%
$ & $~\left(  \frac{1}{2},\frac{1}{2}\right)  $ & $%
\begin{array}
[c]{c}%
A_{\alpha\dot{\beta}}^{h=+1}\left(  \pi,\bar{\pi}\right)  =\frac{w_{\alpha
}\bar{\pi}_{\dot{\beta}}}{\langle\pi w\rangle}\\
A_{\alpha\dot{\beta}}^{h=-1}\left(  \pi,\bar{\pi}\right)  =\frac{\pi_{\alpha
}\bar{w}_{\dot{\beta}}}{\langle\bar{\pi}\bar{w}\rangle}%
\end{array}
$\\\hline
$%
\genfrac{}{}{0pt}{}{\text{field strength}}{F_{\mu\nu}\;\;\;\;\;\;\;\ \;\;}%
$ & $%
\begin{array}
[c]{c}%
(0,1)\\
(1,0)
\end{array}
$ & $%
\begin{array}
[c]{c}%
F_{\dot{\alpha}\dot{\beta}}^{h=+1}\left(  \pi,\bar{\pi}\right)  =\bar{\pi
}_{\acute{\alpha}}\bar{\pi}_{\dot{\beta}}\\
F_{\alpha\beta}^{h=-1}\left(  \pi,\bar{\pi}\right)  =\pi_{\alpha}\pi_{\beta}%
\end{array}
$\\\hline
$%
\genfrac{}{}{0pt}{}{\text{metric}}{g_{\mu\nu}\;\;\;}%
$ & $~\left(  1,1\right)  $ & $%
\begin{array}
[c]{c}%
g_{\dot{\alpha}\dot{\beta}\gamma\delta}^{h=+2}\left(  \pi,\bar{\pi}\right)
=\frac{\bar{\pi}_{\acute{\alpha}}\bar{\pi}_{\dot{\beta}}w_{\gamma}w_{\delta}%
}{\langle\pi w\rangle^{2}}\\
g_{\alpha\beta\dot{\gamma}\dot{\delta}}^{h=-2}\left(  \pi,\bar{\pi}\right)
=\frac{\pi_{\alpha}\pi_{\beta}\bar{w}_{\dot{\gamma}}\bar{w}_{\dot{\delta}}%
}{\langle\bar{\pi}\bar{w}\rangle^{2}}%
\end{array}
$\\\hline
$%
\genfrac{}{}{0pt}{}{\text{curvature}}{R_{\mu\nu\lambda\sigma}\;\;\;\;}%
$ & $%
\begin{array}
[c]{c}%
(0,2)\\
(2,0)
\end{array}
$ & $%
\begin{array}
[c]{c}%
R_{\dot{\alpha}\dot{\beta}\dot{\gamma}\dot{\delta}}^{h=+2}\left(  \pi,\bar
{\pi}\right)  =\bar{\pi}_{\acute{\alpha}}\bar{\pi}_{\dot{\beta}}\bar{\pi
}_{\dot{\gamma}}\bar{\pi}_{\dot{\delta}}\\
R_{\alpha\beta\gamma\delta}^{h=-2}\left(  \pi,\bar{\pi}\right)  =\pi_{\alpha
}\pi_{\beta}\pi_{\gamma}\pi_{\delta}%
\end{array}
$\\\hline\hline
\end{tabular}
\label{Zk}%
\end{equation}
The field strength $F_{\mu\nu}=\partial_{\lbrack\mu}A_{\nu]}$ can be written
in terms of the gauge potential in momentum and spinor space for an arbitrary
combination of both helicities as follows
\begin{align}
A_{\alpha\dot{\beta}}  &  =a^{+}A_{\alpha\dot{\beta}}^{+}\left(  \pi,\bar{\pi
}\right)  +a^{-}A_{\alpha\dot{\beta}}^{-}\left(  \pi,\bar{\pi}\right) \\
F_{\alpha\dot{\beta}\gamma\dot{\delta}}  &  =k_{\alpha\dot{\beta}}%
A_{\gamma\dot{\delta}}-k_{\gamma\dot{\delta}}A_{\alpha\dot{\beta}}%
=\varepsilon_{\alpha\gamma}a^{+}F_{\dot{\beta}\dot{\delta}}^{+}\left(
\pi,\bar{\pi}\right)  +\varepsilon_{\dot{\beta}\dot{\delta}}a^{-}%
F_{\alpha\gamma}^{-}\left(  \pi,\bar{\pi}\right)
\end{align}
which is consistent with the wavefunctions $A^{\pm}\left(  \pi,\bar{\pi
}\right)  ,F^{\pm}\left(  \pi,\bar{\pi}\right)  $ given in the table above.
Note that all wavefunctions are automatically transverse to $k_{\alpha
\dot{\beta}}=\pi_{\alpha}\bar{\pi}_{\dot{\beta}}$ under the Lorentz invariant
dot product using the metric in spinor space $\varepsilon^{\alpha\beta}%
\oplus\varepsilon^{\dot{\alpha}\dot{\beta}}.$

In field theory computations that use twistor techniques \cite{cachazo}, the
twistor space wavefunctions above are used for the corresponding physical
external particles with \textit{definite momentum}, up to overall normalizations.

\section{2T-physics}

As mentioned above, it has been discovered recently that there are many ways
of solving the same constraints on the twistor $Z_{A}$ and derive other
relations between $\mu,\lambda$ and phase space \cite{twistorBP1}. These other
solutions describe not only the massless particle, but also massive particle,
relativistic or non-relativistic, in flat space or curved space, interacting
or non-interacting, as shown in the examples in Fig.1. These new twistors were
discovered by using two time physics (2T-physics) as a technique.

2T-physics was also used to obtain the generalization of twistors to higher
dimensions, to supersymmetry, and to D-branes. In the rest of these lectures I
will first give a brief outline of the main aspects of 2T-physics and then
summarize these new results.%

\begin{center}
\fbox{\includegraphics[
height=295.6875pt,
width=333.3125pt
]%
{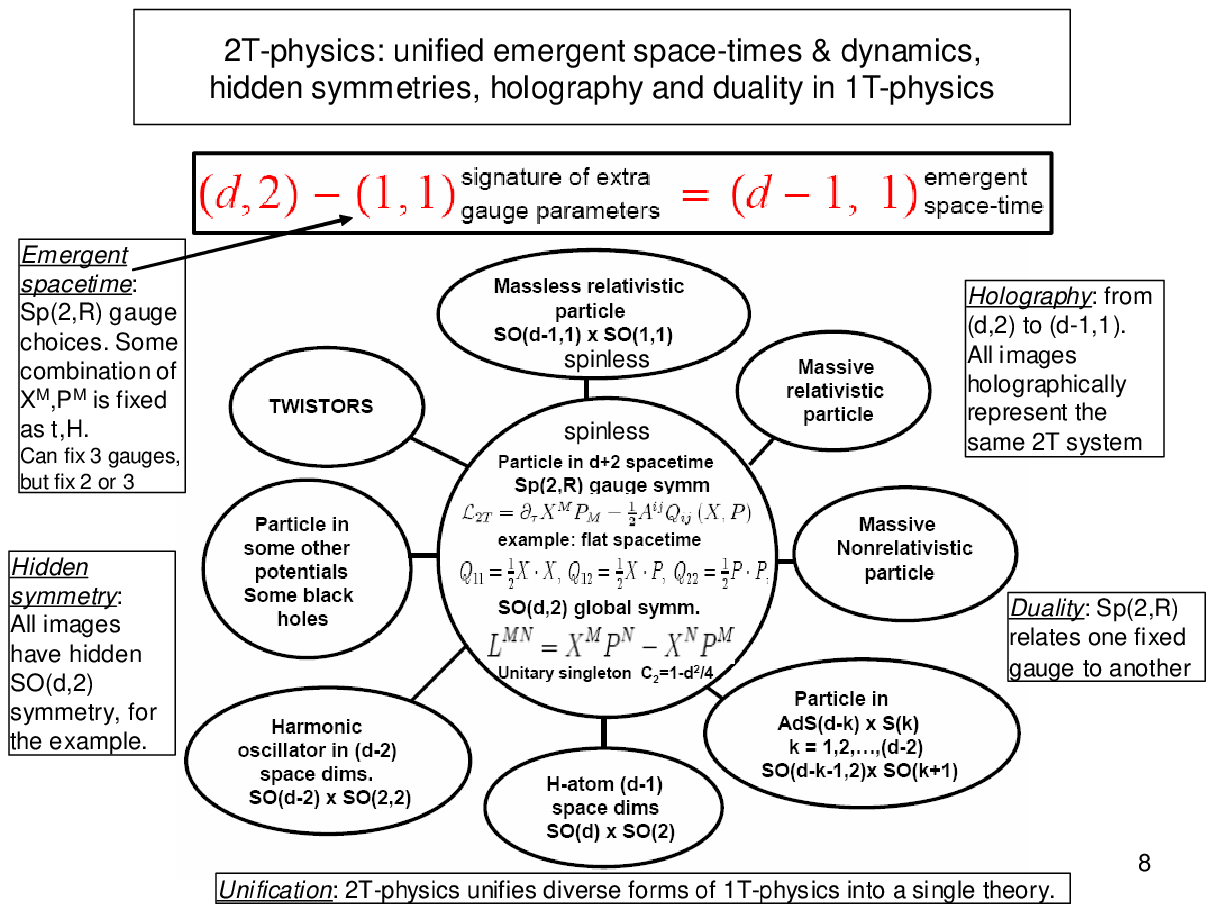}%
}\\
{\small Fig. 1 - 2T-physics in }$d+2${\small \ descends to many 1T-physics
systems in }$\left(  d-1\right)  +1.$%
\label{2tpict}%
\end{center}

\subsection{Emergent spacetimes \& dynamics, holography, duality.}

2T-physics can be viewed as a unification approach for one-time physics
(1T-physics) systems through higher dimensions. It is distinctly different
than Kaluza-Klein theory because there are no Kaluza-Klein towers of states,
but instead there is a family of 1T systems with duality type relationships
among them. The 2T theory is in $d+2$ dimensions, but has enough gauge
symmetry to compensate for the extra $1+1$ dimensions, so that the physical
(gauge invariant) degrees of freedom are equivalent to those encountered in 1T-physics.

One of the strikingly surprising aspects of 2T-physics is that a given $d+2$
dimensional 2T theory descends, through gauge fixing, down to a family of
holographic 1T images in $\left(  d-1\right)  +1$ dimensions. Fig.1 below
illustrates a family of holographic images that have been obtained from the
simplest model of 2T-physics \cite{2tHandAdS}. These include interacting as
well as free systems in 1T-physics.

It must be emphasized that as a by product of the 2T-physics approach certain
physical parameters, such as mass, parameters of spacetime metric, and some
coupling constants appear as moduli in the holographic image while descending
from $d+2$ dimensional phase space to $\left(  d-1\right)  +1$ dimensions or
to twistors.

Each image represented by the ovals around the center in Fig.1 fully captures
the gauge invariant physical content of a unique parent 2T theory that sits at
the center. But from the point of view of 1T-physics each image appears as a
different 1T-dynamical system. The members of such a family naturally must
obey duality-type relationships among them and share many common properties.
In particular they share the same overall global symmetry in $d+2$ dimensions
that becomes hidden and non-linear when acting on the fewer $\left(
d-1\right)  +1$ dimensions in 1T-physics. Thus 2T-physics unifies many 1T
systems into a family that corresponds to a given 2T-physics parent in $d+2$ dimensions.

\subsection{Sp(2,R) gauge symmetry, constraints, solutions and (d,2)}

The essential ingredient in 2T-physics is the basic gauge symmetry Sp(2,R)
acting on phase space $X^{M},P_{M}$ in $d+2$ dimensions. The two timelike
directions is not an input, but is one of the outputs of the Sp$\left(
2,R\right)  $ gauge symmetry. A consequence of this gauge symmetry is that
position and momentum become indistinguishable at any instant, so the symmetry
is of fundamental significance. The transformation of $X^{M},P_{M}$ is
generally a nonlinear map that can be explicitly given in the presence of
background fields \cite{2tbacgrounds}, but in the absence of backgrounds the
transformation reduces to a linear doublet action of Sp$\left(  2,R\right)  $
on $\left(  X^{M},P^{M}\right)  $ for each $M$ \cite{2treviews}. The physical
phase space is the subspace that is gauge invariant under Sp$\left(
2,R\right)  .$ Since Sp$\left(  2,R\right)  $ has 3 generators, to reach the
physical space we must choose 3 gauges and solve 3 constraints. So, the gauge
invariant subspace of $d+2$ dimensional phase space $X^{M},P_{M}$ is a phase
space with six fewer degrees of freedom in $\left(  d-1\right)  $
\textit{space} dimensions $\left(  x^{i},p_{i}\right)  ,$ $i=1,2,\cdots\left(
d-1\right)  .$

In some cases it is more convenient not to fully use the three Sp$\left(
2,R\right)  $ gauge symmetry parameters and work with an intermediate space in
$\left(  d-1\right)  +1$ dimensions $\left(  x^{\mu},p_{\mu}\right)  ,$ that
includes time. This space can be further reduced to $d-1$ space dimensions
$\left(  x^{i},p_{i}\right)  $ by a remaining one-parameter gauge symmetry.

There are many possible ways to embed the $\left(  d-1\right)  +1$ or $\left(
d-1\right)  $ phase space in $d+2$ phase space, and this is done by making
Sp(2,R) gauge choices. In the resulting gauge fixed 1T system, time,
Hamiltonian, and in general curved spacetime, are emergent concepts. The
Hamiltonian, and therefore the dynamics as tracked by the emergent time, may
look quite different in one gauge versus another gauge in terms of the
remaining gauge fixed degrees of freedom. In this way, a unique 2T-physics
action gives rise to many 1T-physics systems.

A particle interacting with various backgrounds in $\left(  d-1\right)  +1$
dimensions (e.g. electromagnetism, gravity, high spin fields, any potential,
etc.), usually described in a worldline formalism in 1T-physics, can be
equivalently described in 2T-physics.

The general 2T theory for a particle moving in any background field has been
constructed \cite{2tbacgrounds}. For a spinless particle it takes the form%
\begin{equation}
S=\int d\tau~\left(  \dot{X}^{iM}P_{M}-\frac{1}{2}A^{ij}Q_{ij}\left(
X,P\right)  \right)  ,
\end{equation}
where the symmetric $A^{ij}\left(  \tau\right)  $ $,$ $i,j=1,2,$ is the
Sp$\left(  2,R\right)  $ gauge field, and the three Sp$\left(  2,R\right)  $
generators $Q_{ij}\left(  X\left(  \tau\right)  ,P\left(  \tau\right)
\right)  ,$ which generally depend on background fields that are functions of
$\left(  X\left(  \tau\right)  ,P\left(  \tau\right)  \right)  $, are required
to form an Sp$\left(  2,R\right)  $ algebra. The background fields must
satisfy certain conditions to comply with the Sp$\left(  2,R\right)  $
requirement. An infinite number of solutions to the requirement can be
constructed \cite{2tbacgrounds}. So any 1T particle worldline theory, with any
backgrounds, can be obtained as a gauge fixed version of some 2T particle
worldline theory.

The 1T systems which appear in the diagram above are obtained by considering
the simplest version of 2T-physics without any background fields. The 2T
action for a \textquotedblleft free\textquotedblright\ 2T particle is
\cite{2treviews}
\begin{equation}
S_{2T}=\frac{1}{2}\int d\tau~D_{\tau}X_{i}^{M}X_{j}^{N}\eta_{MN}%
\varepsilon^{ij}=\int d\tau~\left(  \dot{X}^{M}P^{N}-\frac{1}{2}A^{ij}%
X_{i}^{M}X_{j}^{N}\right)  \eta_{MN}. \label{2Taction}%
\end{equation}
Here $X_{i}^{M}=\left(  X^{M}~P^{M}\right)  ,$ $i=1,2,$ is a doublet under
Sp$\left(  2,R\right)  $ for every $M,$ the structure $D_{\tau}X_{i}%
^{M}=\partial_{\tau}X_{i}^{M}-A_{i}^{~j}X_{j}^{M}$ is the Sp(2,R) gauge
covariant derivative, Sp(2,R) indices are raised and lowered with the
antisymmetric Sp$\left(  2,R\right)  $ metric $\varepsilon^{ij},$ and in the
last expression an irrelevant total derivative $-\left(  1/2\right)
\partial_{\tau}\left(  X\cdot P\right)  $ is dropped from the action. This
action describes a particle that obeys the Sp$(2,R)$ gauge symmetry, so its
momentum and position are locally indistinguishable due to the gauge symmetry.
The $\left(  X^{M},P^{M}\right)  $ satisfy the Sp$\left(  2,R\right)  $
constraints
\begin{equation}
Q_{ij}=X_{i}\cdot X_{j}=0:\;X\cdot X=P\cdot P=X\cdot P=0,
\label{2Tconstraints}%
\end{equation}
that follow from the equations of motion for $A^{ij}$. The vanishing of the
gauge symmetry generators $Q_{ij}=0$ implies that the physical phase space is
the subspace that is Sp$\left(  2,R\right)  $ gauge invariant. These
constraints have non-trivial solutions only if the metric $\eta_{MN}$ has two
timelike dimensions. So when position and momentum are locally
indistinguishable, to have a non-trivial system, two timelike dimensions are
necessary as a consequence of the Sp$\left(  2,R\right)  $ gauge symmetry.

Thus the $\left(  X^{M},P^{M}\right)  $ in Eq.(\ref{2Taction}) are SO$\left(
d,2\right)  $ vectors, labeled by $M=0^{\prime},1^{\prime},\mu$ or
$M=\pm^{\prime},\mu,$ and $\mu=0,1,\cdots,\left(  d-1\right)  $ or $\mu
=\pm,1,\cdots,\left(  d-2\right)  ,$ with lightcone type definitions of
$X^{\pm^{\prime}}=\frac{1}{\sqrt{2}}\left(  X^{0^{\prime}}\pm X^{1^{\prime}%
}\right)  $ and $X^{\pm}=\frac{1}{\sqrt{2}}\left(  X^{0}\pm X^{3}\right)  .$
The SO$\left(  d,2\right)  $ metric $\eta^{MN}$ is given by
\begin{align}
ds^{2}  &  =dX^{M}dX^{N}\eta_{MN}=-2dX^{+^{\prime}}dX^{-^{\prime}}+dX^{\mu
}dX^{\nu}\eta_{\mu\nu}\\
&  =-\left(  dX^{0^{\prime}}\right)  ^{2}+\left(  dX^{1^{\prime}}\right)
^{2}-\left(  dX^{0}\right)  ^{2}+\left(  dX^{1}\right)  ^{2}+\left(
dX_{\perp}\right)  ^{2}\\
&  =-2dX^{+^{\prime}}dX^{-^{\prime}}-2dX^{+}dX^{-}+\left(  dX_{\perp}\right)
^{2}.
\end{align}
where the notation $X_{\perp}$ indicates SO$\left(  d-2\right)  $ vectors.

\subsection{SO(d,2) global symmetry, quantization and the singleton
\label{L2section}}

The target phase space $X^{M},P_{M}$ is flat in $d+2$ dimension, and hence the
system in Eq.(\ref{2Taction}) has an SO$(d,2)$ global symmetry. The conserved
generators of SO$\left(  d,2\right)  $
\begin{equation}
L^{MN}=X^{M}P^{N}-X^{N}P^{M},\;\partial_{\tau}L^{MN}=0,
\end{equation}
commute with the SO$\left(  d,2\right)  $ invariant Sp$\left(  2,R\right)  $
generators $X\cdot X$, $P\cdot P$, $X\cdot P$. It will be useful to consider
the matrix%
\begin{equation}
\left(  L\right)  _{A}^{~B}=\frac{1}{4i}L_{MN}\left(  \Gamma^{MN}\right)
_{A}^{~B}%
\end{equation}
constructed by using the $d$-dimensional analogs of the gamma
matrices in Eqs.(\ref{gam1},\ref{gam2}) (see footnotes 5,6 for
details).

If the square of the matrix $L^{2}$ is computed at the classical level, i.e.
not caring about the orders of generators $L_{MN},$ then one finds that
$\left(  L^{2}\right)  _{A}^{~B}$ is proportional to the identity matrix
$\delta_{A}^{B},$ $\left(  L^{2}\right)  =\left(  \frac{1}{4i}\Gamma
_{MN}L^{MN}\right)  ^{2}=\frac{1}{8}L^{MN}L_{MN}~1.$ Furthermore by computing,
still at the classical level $\frac{1}{2}L^{MN}L_{MN}=X^{2}P^{2}-\left(
X\cdot P\right)  ^{2},$ and imposing the classical constraints $X^{2}%
=P^{2}=\left(  X\cdot P\right)  =0,$ one finds that $L^{2}=0$ in the
space of gauge invariants of the classical theory. By taking higher
powers of $L,$ we find $L^{n}=0$ for all positive integers $n\geq2.$
This is a very special non-trivial representation of the non-compact
group SO$\left( d,2\right)  _{L}$, and all classical gauge
invariants, which are functions of $L^{MN},$ can be classified as
irreducible multiplets of SO$\left(  d,2\right)  _{L}$.

We now consider the SO(d,2) covariant quantization of the theory. In
the quantum theory the $L_{MN}$ form the Lie algebra of SO$\left(
d,2\right) ,$ therefore if the square of the matrix $L$ is computed
at the quantum level, by taking into account the orders
of the operators $L^{MN},$ one finds%
\begin{equation}
L^{2}=\left(  \frac{1}{4i}\Gamma_{MN}L^{MN}\right)
^{2}=-\frac{d}{2}\left( \frac{1}{4i}\Gamma_{MN}L^{MN}\right)
+\frac{1}{8}L^{MN}L_{MN}~1.\label{L2}
\end{equation}
In this computation we used the properties of gamma matrices%
\begin{gather*}
\Gamma_{MN}\Gamma_{RS}=\Gamma_{MNRS}+\left(  \eta_{NR}\eta_{MS}-\eta_{MR}%
\eta_{NS}\right) \\
+\left(  \eta_{NR}\Gamma_{MS}-\eta_{MR}\Gamma_{NS}-\eta_{NS}\Gamma_{MR}%
+\eta_{MS}\Gamma_{NR}\right)  .
\end{gather*}
The term $\Gamma_{MNRS}L^{MN}L^{RS}$ vanishes for $L^{MN}=X^{[M}P^{N]}$ due to
a clash between symmetry/antisymmetry. The term \textquotedblleft$\eta
_{NR}\Gamma_{MS}\cdots$\textquotedblright\ turns into a commutator, and after
using the SO$\left(  d,2\right)  $ Lie algebra for $\left[  L^{MN}%
,L^{RS}\right]  $ it produces the linear term proportional to $d/2$ in
Eq.(\ref{L2}). The term \textquotedblleft$\eta_{NR}\eta_{MS}\cdots
$\textquotedblright\ produces the last term in Eq.(\ref{L2}). Furthermore the
Casimir $\frac{1}{2}L^{MN}L_{MN}$ does not vanish at the quantum level. As
shown in \cite{2treviews}, in the Sp$\left(  2,R\right)  $ gauge invariant
physical sector of phase space one finds that it has the fixed value $\frac
{1}{2}L^{MN}L_{MN}=1-d^{2}/4$ rather than zero. Hence, in the physical sector
of the quantum theory the matrix $L_{A}^{~B}$ satisfies the following algebra%
\begin{equation}
\left(  L^{2}\right)  _{A}^{~B}=-\frac{d}{2}~L_{A}^{~B}+\frac{1}{8}\left(
1-\frac{d^{2}}{4}\right)  ~\delta_{A}^{~B},\;\text{on physical states.}
\label{LLtoL}%
\end{equation}
We compute the higher powers $L^{n}$ on physical states by
repeatedly using this relation, and find a similar form with
constant coefficients that are determined only by $\alpha,\beta$
\begin{equation}
\left(  L^{n}\right)
_{A}^{~B}=\alpha_{n}~L_{A}^{~B}+\beta_{n}~\delta _{A}^{~B}.
\end{equation}
We can then compute the Casimir eigenvalues\footnote{Note that in
the literature one may find that the definition of the cubic and
higher Casimir eigenvalues
are given as a linear combination of our $C_{n}.$} $C_{n}=\frac{1}{s_{d}%
}Tr\left(  \left(  2L\right)  ^{n}\right)  =2^{n}\beta_{n}$. Evidently the
$C_{n}$ will end up having fixed values determined by the dimension $d$ of
SO$\left(  d,2\right)  _{R}.$ In particular,
\begin{equation}
C_{2}=1-\frac{d^{2}}{4},\;\;C_{3}=-d\left(  1-\frac{d^{2}}{4}\right)
,\;C_{4}=\left(  1-\frac{d^{2}}{4}\right)  \left(  1+\frac{3d^{2}}{4}\right)
,\;\text{etc.}%
\end{equation}
Therefore, at the quantum level we have identified a special unitary
representation that classifies all physical states of the theory.
This is the singleton representation of SO$\left(  d,2\right)  $ for
any $d$. Our approach shows that the singleton is more fully
characterized by the constraints satisfied by the charges in
Eq.(\ref{LLtoL}).

\subsection{Twistors for other particle dynamics}

We now introduce the general twistor transform that applies not only to
massless particles, but to other particle systems shown in Fig.1 in any
dimension $d.$ The basic idea \cite{2ttwistor} follows from the formalism in
the following section that includes supersymmetry. Here we give the result for
spinless particles and without supersymmetry in $d$ dimensions
\cite{twistorBP2} and comment on its properties. The general twistor transform
is \cite{twistorBP2}
\begin{equation}
Z=\left(
\genfrac{}{}{0pt}{}{\mu}{\lambda}%
\right)  ,\;\mu=-i\frac{X_{\mu}\bar{\gamma}^{\mu}}{\sqrt{2}X^{+^{\prime}}%
}\lambda,\;\text{and ~}\lambda\bar{\lambda}=\frac{1}{\sqrt{2}}\left(
X^{+^{\prime}}P^{\mu}-P^{+^{\prime}}X^{\mu}\right)  \gamma_{\mu}%
,\;\label{gentwist}%
\end{equation}
Here $Z_{A}^{~a}$ is a $s_{d}\times\frac{s_{d}}{4}$ matrix with $A=1,2,\cdots
,s_{d}$ and $a=1,2,\cdots,\frac{s_{d}}{4},$ and $\mu,\lambda$ are $\frac
{s_{d}}{2}\times\frac{s_{d}}{4}$ matrices, where $s_{d}=2^{d/2}$ is the
dimension of the Weyl spinor for SO$\left(  d,2\right)  $ for even $d$. For
$d=4$ this reduces to the quartet $Z_{A}$ that we discussed in section
(\ref{sectwistors}). The $\gamma^{\mu},\bar{\gamma}^{\mu}$ are the gamma
matrices for even $d$ dimensions in the two Weyl bases (analog of Pauli
matrices in section (\ref{sectwistors}))$.$ The general twistor $Z_{A}^{~a}$
automatically satisfies the following $\left(  \frac{s_{d}}{4}\right)  ^{2}$
constraints by construction%
\begin{equation}
\left(  \bar{Z}Z\right)  _{a}^{~b}=\left(  \bar{\lambda}~\bar{\mu}\right)
\left(
\genfrac{}{}{0pt}{}{\mu}{\lambda}%
\right)  =\left(  \bar{\lambda}\mu+\bar{\mu}\lambda\right)  _{a}^{~b}=0.
\end{equation}
We regard this constraint as the generator of a gauge symmetry that acts on
the $a$ index, and introduce a gauge field $V_{a}^{~b}$ associated with this constraint.

The $A$ index on $Z_{A}^{~a}$ is the basis for the SO$\left(  d,2\right)  $
spinor. This is the global symmetry whose generators $L^{MN}$ can be
constructed from either the twistors $\bar{Z},Z$ or from phase space
$X^{M},P^{M}.$ Indeed the twistor transform above is constructed to satisfy
the relation \cite{twistorBP2}
\begin{equation}
\frac{1}{4i}L^{MN}\left(  \Gamma_{MN}\right)  _{A}^{~B}=L_{A}^{~B}=\left(
Z\bar{Z}\right)  _{A}^{~B}-\frac{1}{s_{d}}Tr\left(  Z\bar{Z}\right)
~\delta_{A}^{~B}.
\end{equation}
The trace term automatically vanishes if $Z$ is constructed to satisfy
$\left(  \bar{Z}Z\right)  _{a}^{~b}=0$ as above.

By inserting the twistor transform into the following twistor action (in which
$\bar{Z}_{b}^{~A}Z_{A}^{~a}$ already vanishes) we derive the phase space
action that determines the canonical structure for the phase space $\left(
X^{M},P^{M}\right)  $ in $d+2$ dimensions%
\begin{align}
S\left(  Z\right)   &  =\frac{4}{s_{d}}\int d\tau~\left(  i\partial_{\tau
}Z_{A}^{~a}\bar{Z}_{a}^{~A}+Z_{A}^{~a}V_{a}^{~b}\bar{Z}_{b}^{~A}\right)  \\
&  =i\frac{4}{s_{d}}\int d\tau~Tr\left(  \partial_{\tau}\mu\bar{\lambda
}+\partial_{\tau}\lambda\bar{\mu}\right)  =\frac{4}{s_{d}}\int d\tau~Tr\left(
\partial_{\tau}\left(  \frac{X_{\mu}\bar{\gamma}^{\mu}}{\sqrt{2}X^{+^{\prime}%
}}\right)  \lambda\bar{\lambda}\right)  \\
&  =\frac{4}{s_{d}}\int d\tau~\frac{1}{\sqrt{2}}\left(  X^{+^{\prime}}P^{\mu
}-P^{+^{\prime}}X^{\mu}\right)  Tr\left(  \partial_{\tau}\left(  \frac{X_{\mu
}\bar{\gamma}^{\mu}}{\sqrt{2}X^{+^{\prime}}}\right)  \gamma_{\mu}\right)  \\
&  =\int d\tau~\left(  X^{+^{\prime}}P_{\mu}-P^{+^{\prime}}X_{\mu}\right)
\partial_{\tau}\left(  \frac{X^{\mu}}{X^{+^{\prime}}}\right)  \\
&  =\int d\tau~\left(  \partial_{\tau}X^{\mu}-\frac{\partial_{\tau
}X^{+^{\prime}}}{X^{+^{\prime}}}X^{\mu}\right)  \left(  P_{\mu}-\frac
{P^{+^{\prime}}}{X^{+^{\prime}}}X_{\mu}\right)  \\
&  =\int d\tau~\left(  \partial_{\tau}X^{\mu}P_{\mu}-\partial_{\tau
}X^{+^{\prime}}P^{-^{\prime}}-\partial_{\tau}X^{-^{\prime}}P^{+^{\prime}%
}\right)  =\int d\tau~\partial_{\tau}X^{M}P_{M}.
\end{align}
The last line follows thanks to the constraints $X^{2}=P^{2}=X\cdot
P=0$ that are satisfied in the Sp$\left(  2,R\right)  $ invariant
physical sector. This shows the consistency of our twistor transform
of Eq.(\ref{gentwist}) for spinless particles in all dimensions.
Hence the 2T-physics system in $d+2$ dimensions is reproduced by the
twistor $Z_{A}^{~a}$ with the given properties.

Now we can choose explicitly the Sp$\left(  2,R\right)  $ gauges that reduce
the 2T-physics system to the the various holographic pictures given in Fig.1.
By inserting the gauge fixed forms of $\left(  X^{M},P^{M}\right)  $ we will
obtain the twistor transforms for all the holographic pictures.

The SO$\left(  d-1,1\right)  $ covariant massless particle emerges if we
choose the two gauges, $X^{+^{\prime}}\left(  \tau\right)  =1$ and
$P^{+^{\prime}}\left(  \tau\right)  =0$, and solve the two constraints
$X^{2}=X\cdot P=0$ to obtain the $\left(  d-1\right)  +1$ dimensional phase
space $\left(  x^{\mu},p_{\mu}\right)  $ embedded in $\left(  d+2\right)  $
dimensions
\begin{align}
X^{M} &  =\left(  \overset{+^{\prime}}{1},\;\overset{-^{\prime}}{x^{2}%
/2}~,\;\overset{\mu}{x^{\mu}}\right)  ,\label{massless1}\\
P^{M} &  =\left(  ~0~,~x\cdot p~,\;~p^{\mu}\right)  .\label{massless2}%
\end{align}
The remaining constraint, $P^{2}=-2P^{+^{\prime}}P^{-^{\prime}}+P^{\mu}P_{\mu
}=p^{2}=0,$ which is the third Sp$\left(  2,R\right)  $ generator, remains to
be imposed on the physical sector. In this gauge the 2T action reduces to the
relativistic massless particle action in Eq.(\ref{cov})
\begin{equation}
S=\int d\tau~\left(  \dot{X}^{M}P^{N}-\frac{1}{2}A^{ij}X_{i}^{M}X_{j}%
^{N}\right)  \eta_{MN}=\int d\tau\left(  \dot{x}^{\mu}p_{\mu}-\frac{1}%
{2}A^{22}p^{2}\right)  .
\end{equation}
Furthermore, the Sp$\left(  2,R\right)  $ gauge invariant $L^{MN}=X^{M}%
P^{N}-X^{N}P^{M}$ take the following nonlinear form
\begin{equation}
L^{\mu\nu}=x^{\mu}p^{\nu}-x^{\nu}p^{\mu},\;L^{+^{\prime}-^{\prime}}=x\cdot
p,\;L^{+^{\prime}\mu}=p^{\mu},\;L^{-^{\prime}\mu}=\frac{x^{2}}{2}p^{\mu
}-x^{\mu}x\cdot p.\label{conf0}%
\end{equation}
These are recognized as the generators of SO$\left(  d,2\right)  $ conformal
transformations of the $\left(  d-1\right)  +1$ dimensional phase space at the
classical level. Thus the hidden conformal symmetry of the massless system is
understood as the Lorentz symmetry in $d+2$ dimensions.

Inserting the gauge fixed form of $\left(  X^{M},P^{M}\right)  $ of
Eqs.(\ref{massless1},\ref{massless2}) into the general twistor transform in
Eq.(\ref{gentwist}), and specializing to four dimensions $d=4,$ we obtain the
Penrose version of twistor transform Eqs.(\ref{pm},\ref{line}) for the
massless spinless particle. Note that for $d\geq4$ the rectangular twistor
$Z_{A}^{~a}$ has several columns, which is a structure that is absent in the
Penrose transform in $d=4.$ The columns of the higher dimensional twistor
satisfy many relations among themselves since they only depend only on the
vectors $X^{M},P^{M}.$

The parent theory can be gauge fixed in many ways that produce not only the
massless particle, but also an assortment of other particle dynamical systems
\cite{2treviews}\cite{2tHandAdS}\cite{twistorBP1}. To emphasize this point we
give also the massive relativistic particle gauge by fixing two gauges and
solving the constraints $X^{2}=X\cdot P=0$ explicitly as follows
\begin{align}
X^{M}  &  =\left(  \overset{+^{\prime}}{\frac{1+a\mathbf{\;}}{2a\mathbf{\;}}%
},\;\overset{-^{\prime}}{\;\frac{x^{2}a\mathbf{\;}}{1+a\mathbf{\;}}%
}~,~~\overset{\mu}{x^{\mu}}\right)  ,\;a\equiv\sqrt{1+\frac{m^{2}x^{2}%
}{\left(  x\cdot p\right)  ^{2}}}\label{massive2x}\\
P^{M}  &  =\left(  \frac{-m^{2}}{2(x\cdot p)a},\;\;\left(  x\cdot p\right)
a\;\mathbf{,\;\;}p^{\mu}\right)  ,\;P^{2}=p^{2}+m^{2}=0. \label{massive2p}%
\end{align}
In this gauge the 2T action reduces to the relativistic massive particle
action
\begin{equation}
S=\int d\tau~\left(  \dot{X}^{M}P^{N}-\frac{1}{2}A^{ij}X_{i}^{M}X_{j}%
^{N}\right)  \eta_{MN}=\int d\tau\left(  \dot{x}^{\mu}p_{\mu}-\frac{1}%
{2}A^{22}\left(  p^{2}+m^{2}\right)  \right)  .
\end{equation}
The twistor for the massive particle follows from Eqs.(\ref{massive2x}%
,\ref{massive2p},\ref{gentwist}) as shown in \cite{twistorBP1}%
\begin{equation}
\mu^{\dot{\alpha}}=-ix^{\dot{\alpha}\beta}\lambda_{\beta}\frac{2a}%
{1+a},\;\;\lambda_{\alpha}\bar{\lambda}_{\dot{\beta}}=\frac{1+a\mathbf{\;}%
}{2a\mathbf{\;}}p_{\alpha\dot{\beta}}+\frac{m^{2}}{2(x\cdot p)a}x_{\alpha
\dot{\beta}}. \label{twistormassive}%
\end{equation}

A little recognized fact is that this action is invariant under SO$\left(
d,2\right)  $. This SO$\left(  d,2\right)  $ does not have the form of
conformal transformations of Eq.(\ref{conf0}), but is a deformed version of
it, including the mass parameter. Its generators are obtained by inserting the
massive particle gauge into the gauge invariant $L^{MN}=X^{M}P^{N}-X^{N}P^{M}$%
\begin{align}
L^{\mu\nu}  &  =x^{\mu}p^{\nu}-x^{\nu}p^{\mu},\text{ \ \ \ }L^{+^{\prime
}-^{\prime}}=\left(  x\cdot p\right)  a,\label{LMNmassive1}\\
L^{+^{\prime}\mu}  &  =\frac{1+a\mathbf{\;}}{2a\mathbf{\;}}p^{\mu}+\frac
{m^{2}}{2\left(  x\cdot p\right)  a}x^{\mu}\label{LMNmassive2}\\
L^{-^{\prime}\mu}  &  =\frac{x^{2}a\mathbf{\;}}{1+a\mathbf{\;}}p^{\mu}-\left(
x\cdot p\right)  ax^{\mu} \label{LMNmassive3}%
\end{align}
It can be checked explicitly that the massive particle action above is
invariant under the SO$\left(  d,2\right)  $ transformations generated by the
Poisson brackets $\delta x^{\mu}=\frac{1}{2}\omega_{MN}\left\{  L^{MN},x^{\mu
}\right\}  $ and $\delta p^{\mu}=\frac{1}{2}\omega_{MN}\left\{  L^{MN},p^{\mu
}\right\}  ,$ up to a reparametrization of $A^{22}$ by a scale and an
irrelevant total derivative.

Since both the massive and massless particles give bases for the same
representation of SO$\left(  d,2\right)  $, we must expect a duality
transformation between them. Of course this transformation must be an
Sp$\left(  2,R\right)  =$SL$\left(  2,R\right)  $ local gauge transformation
$\left(
\genfrac{}{}{0pt}{}{\alpha}{\gamma}%
\genfrac{}{}{0pt}{}{\beta}{\delta}%
\right)  \left(  \tau\right)  $ with unit determinant $\alpha\delta
-\beta\gamma=1,$ that transforms the doublets $\left(
\genfrac{}{}{0pt}{}{X^{M}}{P^{M}}%
\right)  \left(  \tau\right)  $ from
Eqs.(\ref{massive2x},\ref{massive2p}) to
Eqs.(\ref{massless1},\ref{massless2}). The
$\alpha,\beta,\gamma,\delta$ are fixed by focussing on the doublets
labeled by $M=+^{\prime}$
\begin{equation}
\left(
\begin{array}
[c]{c}%
\left(  \frac{1+a\mathbf{\;}}{2a\mathbf{\;}}\right) \\
\left(  \frac{-m^{2}}{2(x\cdot p)a}\right)
\end{array}
\right)  =\left(
\begin{array}
[c]{cc}%
\left(  \frac{1+a\mathbf{\;}}{2a\mathbf{\;}}\right)  & 0\\
\left(  \frac{-m^{2}}{2(x\cdot p)a}\right)  & \left(  \frac{2a\mathbf{\;}%
}{1+a\mathbf{\;}}\right)
\end{array}
\right)  \left(
\begin{array}
[c]{c}%
1\\
0
\end{array}
\right)  .
\end{equation}
Applying the inverse of this transformation on the doublets labeled by
$M=\mu$ gives the massless particle phase space (re-labeled by $\left(
\tilde{x}^{\mu},\tilde{p}^{\mu}\right)  $ below) in terms of the massive
particle phase space (labeled by $\left(  x^{\mu},p^{\mu}\right)  $)%
\begin{equation}
\left(
\begin{array}
[c]{cc}%
\left(  \frac{2a\mathbf{\;}}{1+a\mathbf{\;}}\right)  & 0\\
\left(  \frac{m^{2}}{2(x\cdot p)a}\right)  & \left(  \frac{1+a\mathbf{\;}%
}{2a\mathbf{\;}}\right)
\end{array}
\right)  \left(
\begin{array}
[c]{c}%
x^{\mu}\\
p^{\mu}%
\end{array}
\right)  =\left(
\begin{array}
[c]{c}%
\frac{2a\mathbf{\;}}{1+a\mathbf{\;}}x^{\mu}\\
\frac{1+a\mathbf{\;}}{2a\mathbf{\;}}p^{\mu}+\frac{m^{2}}{2(x\cdot p)a}x^{\mu}%
\end{array}
\right)  \equiv\left(
\begin{array}
[c]{c}%
\tilde{x}^{\mu}\\
\tilde{p}^{\mu}%
\end{array}
\right)
\end{equation}
This duality transformation is a canonical transformation $\left\{  \tilde
{x}^{\mu},\tilde{p}^{\nu}\right\}  =\eta^{\mu\nu}=\left\{  x^{\mu},p^{\nu
}\right\}  .$ Also note that the time coordinate $\tilde{x}^{0}$ is different
than the time coordinate $x^{0},$ and so are the corresponding Hamiltonians
for the massless particle $\tilde{p}^{0}=\sqrt{\tilde{p}^{i}\tilde{p}^{i}}$
versus the massive particle $p^{0}=\sqrt{p^{i}p^{i}+m^{2}}.$

The same reasoning applies among all gauge choices of the 2T theory
in Eq.(\ref{2Taction}). All resulting 1T dynamical systems are
holographic images of the same parent theory. Some of the images are
illustrated in the diagram above. In the quantum theory we have
already shown by covariant quantization that the global symmetry
SO$\left(  d,2\right)  $ of the 2T-physics action is realized in the
singleton representation. All the emergent lower dimensional
theories obtained by different forms of gauge fixing, either in the
form of twistors, or in the form of phase space, must also be
realized in the same singleton representation\footnote{At the
classical level all Casimir eigenvalues vanish for the various form
of the SO$\left(  d,2\right)  $ generators. But at the quantum
level, due to ordering of factors that are needed for the correct
closure of the algebra, the Casimir eigenvalues are non-zero and
agree with Eq.(\ref{casimirs}), the singleton representation. The
ordering of the quantum factors has been explicitly performed in the
majority of the holographic images given in Fig.1
\cite{2treviews}\cite{2tHandAdS}.}.

\section{Supersymmetric 2T-physics and twistor gauge}

The 2T-physics action (\ref{2Taction}) and the twistor action
(\ref{action}) are related to one another and can both be obtained
as gauge choices from the same theory in the 2T-physics formalism.
This formalism was introduced in \cite{2ttwistor} and developed
further in the context of the twistor superstring
\cite{2tsuperstring}\cite{2tstringtwistors} and to derive the
general twistor transform \cite{twistorBP1}\cite{twistorBP2}.

\subsection{Coupling $X,P,g$, gauge symmetries, global symmetries.}

In addition to the phase space SO$\left(  d,2\right)  $ vectors $\left(
X^{M},P^{M}\right)  \left(  \tau\right)  ,$ we introduce a group element
$g\left(  \tau\right)  .$ The group $G$ is chosen so that it contains
SO$\left(  d,2\right)  $ as a subgroup in the \textit{spinor} representation.
The simplest case is $G=$SO$\left(  d,2\right)  ,$ and in that case
\begin{equation}
g\left(  \tau\right)  =\exp\left(  \frac{1}{4}\Gamma^{MN}\omega_{MN}\left(
\tau\right)  \right)  \label{g}%
\end{equation}
where $\Gamma^{MN}$ are the gamma matrices\footnote{The trace in spinor space
gives the dimension of the spinor $Tr\left(  1\right)  =s_{d}$ and $Tr\left(
\Gamma^{M}\bar{\Gamma}^{N}\right)  =s_{d}\eta^{MN}.$ For even dimensions
$s_{d}=2^{d/2}$ for the Weyl spinor of SO$\left(  d,2\right)  ,$ and the
$\bar{\Gamma}^{M},\Gamma^{M}$ are the gamma matrices in the bases of the two
different spinor representations$.$The correctly normalized generators of
SO$\left(  d,2\right)  $ in the spinor representation are $S^{MN}=\frac{1}%
{2i}\Gamma^{MN}$, where the gamma matrices satisfy $\Gamma^{M}\bar{\Gamma}%
^{N}+\Gamma^{N}\bar{\Gamma}^{M}=2\eta^{MN},$ while $\Gamma^{MN}=\frac{1}%
{2}\left(  \Gamma^{M}\bar{\Gamma}^{N}-\Gamma^{N}\bar{\Gamma}^{M}\right)  $,
$\Gamma^{MNK}=\frac{1}{3!}\left(  \Gamma^{M}\bar{\Gamma}^{N}\Gamma^{K}%
\mp\text{permutations}\right)  $, etc. There exists a metric $C$ of SO$\left(
d,2\right)  $ in the spinor representation such that when combined with
hermitian conjugation it gives $C^{-1}\left(  \Gamma^{M}\right)  ^{\dagger
}C=-\bar{\Gamma}^{M}$ and $C^{-1}\left(  \Gamma^{MN}\right)  ^{\dagger
}C=-\Gamma^{MN}.$ So the inverse $g^{-1}$ is obtained by combining hermitian
and $C$-conjugation $g^{-1}=C^{-1}\left(  g\right)  ^{\dagger}C\equiv\bar{g}.$
In odd number of dimensions the even-dimension gamma matrices above are
combined to a larger matrix $\hat{\Gamma}^{M}=\left(
\genfrac{}{}{0pt}{}{0}{\Gamma^{M}}%
\genfrac{}{}{0pt}{}{\bar{\Gamma}^{M}}{0}%
\right)  $ for $M=0^{\prime},1^{\prime},0,1,\cdots,\left(  d-2\right)  $ and
add one more matrix for the additional last dimension $\hat{\Gamma}%
^{d-1}=\left(
\genfrac{}{}{0pt}{}{1}{0}%
\genfrac{}{}{0pt}{}{0}{-1}%
\right)  $. The text is written as if $d$ is even; for odd
dimensions we replace everywhere $\hat{\Gamma}^{M}$ for both
$\Gamma^{M}$ and $\bar{\Gamma }^{M}.$ \label{ginverse}} for
SO$\left(  d,2\right)  .$ Table 1 shows the smallest possible
bosonic groups $G$ that contain SO$\left( d,2\right)  $ in the
spinor representation for various dimensions $3\leq d\leq12.$

The table lists all the generators of $G$ as represented by
antisymmetrized
products of gamma matrices $\Gamma^{M_{1}\cdots M_{n}}\equiv\frac{1}%
{n!}\left(  \Gamma^{M_{1}}\bar{\Gamma}^{M_{2}}\Gamma^{M_{3}}\cdots
\Gamma^{M_{n}}\mp\text{permutations}\right)  .$ The criterion for
choosing $G$ is that $G$ is the smallest group whose smallest
fundamental representation has the same dimension $s_{d}$ as the
spinor of SO($d,2).$ Then $\Gamma^{MN}$ (i.e. SO($d,2)$ generators
in the spinor basis) must be included among the generators of $G$.
The number of generators represented by $\Gamma ^{M_{1}\cdots
M_{n}}$ in $d+2$ dimensions is indicated as the subscript. This
number is given by the binomial coefficient $\frac{\left(
d+2\right) !}{n!\left(  d+2-n\right)  !}$ in general, but is divided
by 2 for the case of $\Gamma_{462+}^{{\small M}_{1}{\small \cdots
M}_{6}}$ because this one is self dual for $d+2=12$. The total
number of gamma matrices listed is equal to the number of generators
in $G.$ Taken together these form the Lie algebra of $G$ under
matrix commutation. The following column gives information on
whether the gamma matrices occur in the symmetric or antisymmetric
products of the spinors of SO$\left(  d,2\right)  $, when both
spinor indices $A,B$ are lowered or raised in the form $\left(
\Gamma^{M_{1}\cdots M_{n}}\right) _{AB}$ by using the metric $C$ in
spinor space.

The subset of gamma matrices $\Gamma^{MN}$ represent the SO$\left(
d,2\right)  $ subgroup in $G.$ The gamma matrices ${\small \Gamma}%
^{{\small M}_{1}\cdots{\small M}_{n}}$ with $n\neq2$ lead to degrees
of freedom in the model that correspond to D-branes as explained in
\cite{twistorBP2}. Only the cases of $d=3,4,5,6$ can be constructed
purely with particle degrees of freedom without any D-branes.

\begin{gather*}%
\begin{tabular}
[c]{|l|l|l|l|l|l|}\hline
d & $%
\genfrac{}{}{0pt}{}{\text{SO(d,2})}{\underset{{\large s}_{d}}{\text{spinor}%
}\text{~}}%
$ & G & $\text{G in spin(d,2) basis}$ & $%
\genfrac{}{}{0pt}{}{{\small \Gamma}^{{\small M}_{1}\cdots{\small M}_{n}%
}}{\underset{{\Large s}_{d}\times s_{d}}{\text{in ~product}}}%
$ & G$_{\text{super}}$\\\hline
3 & 4 & Sp$\left(  4,R\right)  $ & ${\small \Gamma}_{10}^{{\small MN}}$ &
$\left(  {\small 4\times4}\right)  _{s}$ & {\small OSp}$\left(  {\small N|4}%
\right)  $\\\hline
4 & 4$,\bar{4}$ & SU$\left(  2,2\right)  $ & ${\small \Gamma}_{15}%
^{{\small MN}}$ & ${\small 4\times\bar{4}}$ & {\small SU}$\left(
{\small 2,2|N}\right)  $\\\hline
5 & 8 & $%
\genfrac{}{}{0pt}{}{\text{spin}^{\ast}\text{(7)~}}{\text{SO}^{\ast}\left(
8\right)  ~~}%
$ & $%
\genfrac{}{}{0pt}{}{\Gamma_{21}^{MN}\text{~~~~~~~~~~~~}}{\Gamma_{21}%
^{MN}\oplus~~\Gamma_{7}^{M}~}%
$ & $\left(  {\small 8\times8}\right)  _{a}$ & $%
\genfrac{}{}{0pt}{}{\text{F(4)~~~~~~~~~~~~~}}{\text{OSp}\left(  8|2N\right)
~}%
$\\\hline
6 & 8$_{+}$ & SO$^{\ast}\left(  8\right)  $ & ${\small \Gamma}_{28}%
^{{\small MN}}$ & $\left(  {\small 8\times8}\right)  _{a}$ & {\small OSp}%
$\left(  {\small 8|2N}\right)  $\\\hline
7 & 16 & SO$^{\ast}\left(  16\right)  $ & ${\small \Gamma}_{36}^{{\small MN}%
}{\small \oplus\Gamma}_{84}^{{\small MNK}}$ & $\left(  1{\small 6\times
16}\right)  _{a}$ & {\small OSp}$\left(  {\small 16|2N}\right)  $\\\hline
8 & {\small 16}$,\overline{{\small 16}}$ & SU$^{\ast}\left(  16\right)  $ &
${\small \Gamma}_{45}^{{\small MN}}{\small \oplus\Gamma}_{210}^{{\small MNKL}%
}$ & ${\small 16\times~}\overline{{\small 16}}$ & {\small SU}$\left(
{\small 16|N}\right)  $\\\hline
9 & 32 & Sp$^{\ast}\left(  32\right)  $ & ${\small \Gamma}_{55}^{{\small MN}%
}{\small \oplus\Gamma}_{11}^{{\small M}}{\small \oplus\Gamma}_{462}%
^{{\small M}_{1}\cdots{\small M}_{5}}$ & $\left(  {\small 32\times32}\right)
_{s}$ & {\small OSp}$\left(  {\small N|32}\right)  $\\\hline
10 & 32$_{+}$ & Sp$^{\ast}\left(  32\right)  $ & ${\small \Gamma}%
_{66}^{{\small MN}}{\small \oplus}\Gamma_{462+}^{{\small M}_{1}{\small \cdots
M}_{6}}$ & $\left(  {\small 32\times32}\right)  _{s}$ & {\small OSp}$\left(
{\small N|32}\right)  $\\\hline
11 & 64 & Sp$^{\ast}\left(  64\right)  $ & ${\small \Gamma}_{78}^{{\small MN}%
}{\small \oplus\Gamma}_{286}^{{\small MNK}}{\small \oplus\Gamma}%
_{1716}^{{\small M}_{1}\cdots{\small M}_{6}}$ & $\left(  {\small 64\times
64}\right)  _{s}$ & {\small OSp}$\left(  {\small N|64}\right)  $\\\hline
12 & {\small 64,}$\overline{{\small 64}}$ & SU$^{\ast}\left(  64\right)  $ &
${\small \Gamma}_{91}^{{\small MN}}{\small \oplus\Gamma}_{1001}^{{\small MNKL}%
}{\small \oplus\Gamma}_{3003}^{{\small M}_{1}\cdots{\small M}_{6}}$ &
${\small 64\times~}\overline{{\small 64}}$ & {\small SU}$\left(
{\small 64|N}\right)  $\\\hline
\end{tabular}
\\
\text{{\small Table 1:} {\small Smallest} {\small group} }{\small G}\text{
{\small that contains Spin}}\left(  {\small d,2}\right)  ;\text{supergroups
}{\small G}_{\text{super}}\text{ }{\small ;}\text{ {\small D-branes}.}%
\end{gather*}

Groups that are larger than the listed $G$ may also be considered in our
scheme in every dimension (e.g. SU$\left(  8\right)  $ instead of SO$\left(
8\right)  $ in $d=6$). In that case the number of generators $\Gamma
^{M_{1}\cdots M_{n}}$ increases compared to the ones listed in the table for
each $d.$ Furthermore the corresponding D-brane degrees of freedom also get
included in the model.

The last column of the table lists the smallest supergroups
$G_{\text{super}}$ that contain $G.$ The number of supersymmetries
depend on the value of $N=0,1,2,\cdots.$ For $N=0$ we just have $G$.
In fact for $N=0,$ the smallest group is SO$\left(  d,2\right)  $
itself for every $d,$ and this would include only particle degrees
of freedom without D-branes for every $d$. The $N=0$ case for either
SO$\left(  d,2\right)  $ or $G$ is discussed in detail in
\cite{twistorBP1}\cite{twistorBP2}.

For $N\neq0,$ the supergroup listed is the smallest supergroup that contains
SO$\left(  d,2\right)  $ in the spinor representation. For physical purposes
the total number of real fermionic generators in $G_{\text{super}}$ cannot
exceed 64 (32 ordinary supercharges and 32 conformal supercharges). For
example, for $d=4$ we can go as far as $N=8,$ since $G_{\text{super}}%
=$SU$\left(  2,2|8\right)  $ has 64 real fermionic parameters. Similarly for
$d=11, $ we cannot have more than $N=1,$ hence OSp$\left(  1|64\right)  $. For
a given $N,$ the $\dot{N}$-dependent bosonic subgroup is then the $R$-symmetry
group that acts on the supercharges. Thus, for $d=4$ and $N=4,$ the
$R$-symmetry subgroup of PSU$\left(  2,2|4\right)  $ is SU$\left(  4\right)
.$

The supergroup element $g\left(  \tau\right)  $ can be written by
exponentiating the Lie superalgebra in the form%
\begin{equation}
\overset{\text{for }d\leq6\text{ the spacetime part of }g\text{ has only
SO(d,2) degrees of freedom }}{\underset{\text{for }d\geq7\text{ supergroups
contain more }\Gamma^{M_{1}\cdots M_{n}},\;\text{which give D-branes~~~}%
}{\text{{\ g}}\left(  \tau\right)  =\exp\left(
\begin{tabular}
[c]{|l|l|}\hline
$\underset{\text{Spin}\left(  d,2\right)  ~\text{subgroup +}\cdots}{\frac
{1}{4}\Gamma^{MN}\omega_{MN}+\cdots}$ & $\underset{\text{fermi}~}%
{\Theta_{{spinor}}^{{i=1\cdots N}}}$\\\hline
$\underset{\text{fermi}~}{\bar{\Theta}}$ & $\underset{~\text{R-symmetry}%
}{R^{a}\omega_{a}}$\\\hline
\end{tabular}
\ \ \right)  }}%
\end{equation}
where $\cdots$ stands for the contributions of the $\Gamma^{M_{1}\cdots M_{n}%
}$ in Table 1, while the $R^{a}$ are the generators of the R-symmetry group.
The generalized 2T superparticle action
\begin{equation}
S_{2T}\left(  X,P,g\right)  =\int d\tau\left[  \frac{1}{2}\varepsilon
^{ij}\partial_{\tau}X_{i}\cdot X_{j}-\frac{1}{2}A^{ij}X_{i}\cdot X_{j}%
+\frac{4}{s_{d}}Str\left(  ig^{-1}\partial_{\tau}gL\right)  \right]  \;\;
\label{S2T}%
\end{equation}
includes the degrees of freedom $\left(  X^{M},P^{M}\right)  $ and those of
the supergroup $g,$ namely $\omega_{MN},\cdots,\omega_{M_{1}\cdots M_{n}}, $
$\omega_{a},$ and $\Theta_{{spinor}}^{{i=1\cdots N}}$. Here the matrix $L $
has the following form%
\begin{equation}
L=\frac{1}{4i}\left(
\genfrac{}{}{0pt}{}{\Gamma^{MN}}{0}%
\genfrac{}{}{0pt}{}{0}{0}%
\right)  L_{MN}=\frac{1}{4i}\left(
\genfrac{}{}{0pt}{}{\left(  \Gamma\cdot X~\bar{\Gamma}\cdot P-\Gamma\cdot
P~\bar{\Gamma}\cdot X\right)  }{0}%
\genfrac{}{}{0pt}{}{0}{0}%
\right)  \label{L}%
\end{equation}

It has been established \cite{2ttwistor} that for $d=3,4,5,6$ and any $N,$
this action descends to the usual superparticle action in $d=3,4,5,6$
dimensions by using the massless particle gauge in Eq.(\ref{massless1}%
,\ref{massless2})
\begin{gather}
S_{2T}^{\text{gauge fixed}}\left(  X,P,g\right)  =S_{d=3,4,5,6}%
^{\text{superparticle}}\left(  x,p,\theta\right) \nonumber\\
=\int d\tau\left[  \dot{x}\cdot p+i\bar{\theta}^{i}\gamma^{\mu}\dot{\theta
}^{i}p_{\mu}-\frac{1}{2}ep^{2}\right]  ,\;i=1,2,\cdots,N. \label{sparticle}%
\end{gather}
So, in this gauge, the supergroups $G_{\text{super}}$ given in the table
above, namely OSp$\left(  N|4\right)  $ in $d=3,$ SU$\left(  2,2|N\right)  $
in $d=4,$ F$\left(  4\right)  $ in $d=5$ and OSp$\left(  8|2N\right)  $ in
$d=6,$ are interpreted as the hidden superconformal symmetries of the
superparticle action \cite{2ttwistor}. The fermions $\theta$ are half of the
fermions $\Theta$ that appear in $g.$ This action has a remaining kappa
supersymmetry and can remove half of the $\theta,$ so only $1/4$ of the
original fermions $\Theta$ are physical.

In $d\geq7$ the action $S_{2T}\left(  X,P,g\right)  $ has D-brane degrees of
freedom in addition to the particle (D0) degrees of freedom \cite{twistorBP2}.
In particular in $d=11$ the brane degrees of freedom are the D2-brane and
D5-brane of M-theory \cite{2ttoyM}.

This action has several local and global symmetries given by \cite{2ttwistor}%
\begin{equation}
\text{local: Sp}\left(  2,R\right)  \times\left(
\genfrac{}{}{0pt}{}{\text{ SO}\left(  d,2\right)  }{\frac{3}{4}\kappa appa}%
\genfrac{}{}{0pt}{}{\frac{3}{4}\kappa appa}{\text{R-symm}}%
\right)  ,\;\text{global: }G_{\text{super}} \label{localglobal1}%
\end{equation}
The global symmetry $G_{\text{super}}$ is evident when $g\left(  \tau\right)
$ is transformed from the left side as $g\rightarrow g^{\prime}\left(
\tau\right)  =g_{L}g\left(  \tau\right)  $ with a $g_{L}\in G_{\text{super}}$
that is $\tau$ independent. Then the Cartan connection $g^{-1}\partial_{\tau
}g$ remains invariant. Noether's theorem gives the conserved $G_{\text{super}%
}$ charges in the form
\begin{equation}
\text{global: }G_{\text{super}}:\text{ }J_{A}^{~B}=\left(  gLg^{-1}\right)
_{A}^{~B}=\left(
\genfrac{}{}{0pt}{}{\text{ G~~~~}}{\text{super}}%
\genfrac{}{}{0pt}{}{\text{super}}{~~~~~\text{R-symm}}%
\right)  _{A}^{~B} \label{J}%
\end{equation}

The local symmetries that act on the \textit{right} side of $g$ have
the matrix form in Eq.(\ref{localglobal1}) which is different than
the matrix form in Eq.(\ref{J}) for the global symmetries that act
on the \textit{left} side of $g.$ Before giving details in the next
paragraph, we mention that , by $\frac{3}{4}\kappa appa$ we imply
that the local kappa supersymmetry can remove as much as $3/4$ of
the fermions in $g.$ The SO$\left(  d,2\right)  $ local symmetry can
remove the SO$\left(  d,2\right)  $ parameters $\omega_{MN} $ in $g$
but cannot remove the additional parameters in $G$ associated with
the other generators $\Gamma^{M_{1}\cdots M_{n}}$ listed in Table 1.
The local R-symmetry can remove from $g$ all of the subgroup
parameters $\omega_{a}$. Thus for $d=3,4,5,6$ cases listed in Table
1, all the bosons in $g$ can be eliminated and $3/4$ of the fermions
can be eliminated, if we wish to choose such a gauge. For $d\geq7$
some of the bosonic degrees of freedom in $g$ cannot be eliminated,
those are related to D-branes.

As in the purely bosonic theory, the local Sp$\left(  2,R\right)
\times\left(
\genfrac{}{}{0pt}{}{\text{ SO}\left(  d,2\right)  }{\frac{3}{4}\kappa appa}%
\genfrac{}{}{0pt}{}{\frac{3}{4}\kappa appa}{\text{R-symm}}%
\right)  $ symmetries can be gauge fixed in a variety of ways to descend to
supersymmetric 1T-systems that are dual to each other, and holographically
represent the same 2T superparticle. The hidden symmetries of any holographic
image is the original global symmetry $G_{\text{super}}.$ Some of these gauge
choices are outlined in the following figure%

\begin{center}
\includegraphics[
height=251.8125pt,
width=335.1875pt
]%
{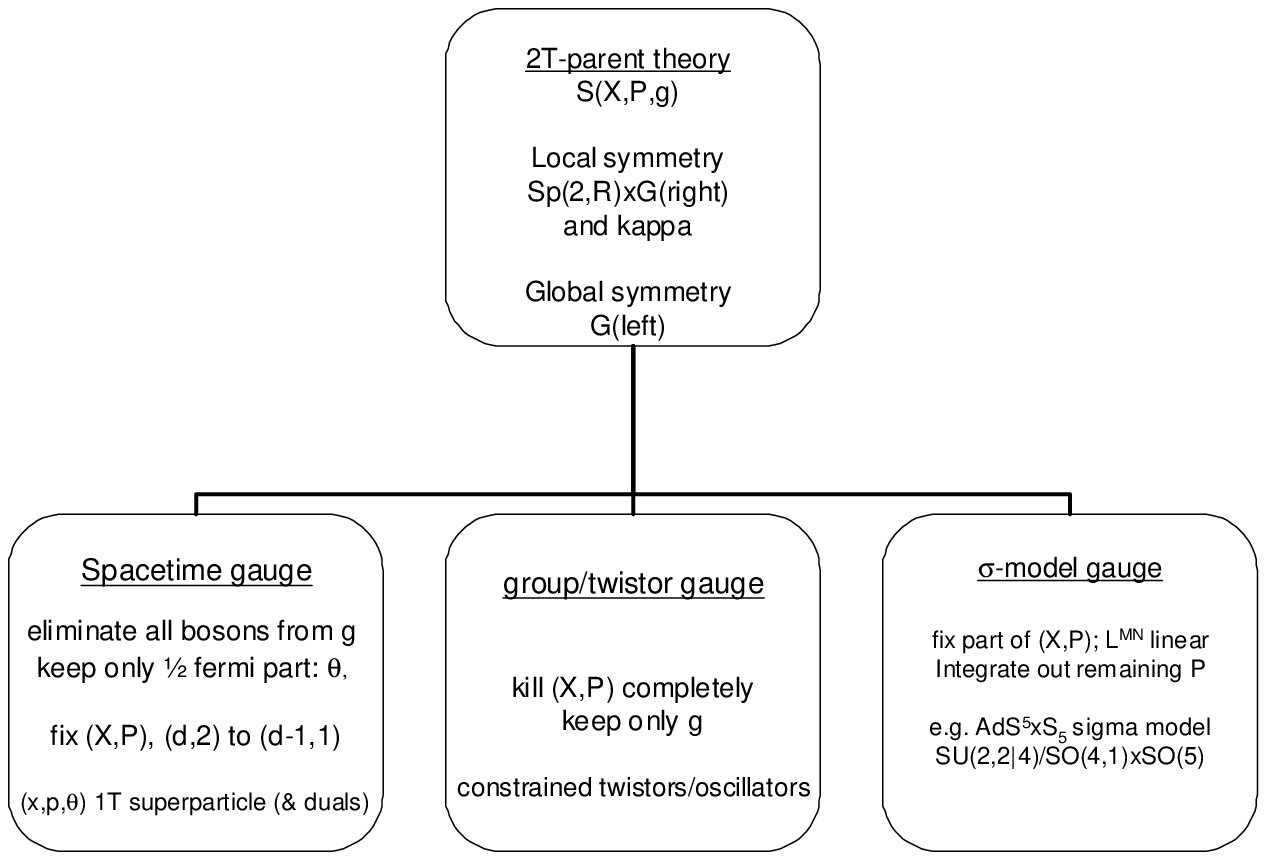}%
\\
Fig. 2 - More dualities, holographic images of 2T superparticle.
\label{fig2dualities}%
\end{center}

Let's outline the properties of the local symmetries Sp$\left(  2,R\right)
\times\left(
\genfrac{}{}{0pt}{}{\text{ SO}\left(  d,2\right)  }{\frac{3}{4}\kappa appa}%
\genfrac{}{}{0pt}{}{~~\frac{3}{4}\kappa appa}{~~\text{R-symm}}%
\right)  $. The local symmetry Sp$\left(  2,R\right)  $ is straightforward
since the first two terms of the action $S_{2T}\left(  X,P,g\right)  $ are the
same as Eq.(\ref{2Taction}). These terms are invariant under Sp$\left(
2,R\right)  $ which acts on $X_{i}^{M}=\left(  X^{M},P^{M}\right)  $ as a
doublet for every $M,$ and on $A^{ij}$ as the gauge field$.$ Furthermore, by
taking $g\left(  \tau\right)  $ as a Sp$\left(  2,R\right)  $ singlet while
noting that $L^{MN}=\varepsilon^{ij}X_{i}^{M}X_{j}^{N}=X^{M}P^{N}-X^{N}P^{M}$
is Sp$\left(  2,R\right)  $ gauge invariant$,$ we see that the full action is
gauge invariant under Sp$\left(  2,R\right)  .$ To see the local symmetry
under SO$\left(  d,2\right)  \times($R-symm) it is convenient to rewrite the
action in the form%
\begin{equation}
S_{2T}\left(  X,P,g\right)  =\int d\tau\left\{
\begin{array}
[c]{c}%
\frac{1}{2s_{d}}\varepsilon^{ij}Str\left[  \partial_{\tau}\left(  g\left(
\genfrac{}{}{0pt}{}{\text{ }X_{i}\cdot\Gamma}{0}%
\genfrac{}{}{0pt}{}{0}{0}%
\right)  g^{-1}\right)  g\left(
\genfrac{}{}{0pt}{}{\text{ }X_{j}\cdot\bar{\Gamma}}{0}%
\genfrac{}{}{0pt}{}{0}{0}%
\right)  g^{-1}\right] \\
-\frac{1}{2}A^{ij}X_{i}\cdot X_{j}%
\end{array}
\right\}  .
\end{equation}
When both $X_{i}^{M}$ and $g\left(  \tau\right)  $ are transformed under
\textit{local} SO$\left(  d,2\right)  \times$R-symm transformations as
$\delta_{R}X_{i}^{M}=\varepsilon_{R}^{MN}X_{iN}$ and $\delta_{R}g=-\frac{1}%
{4}g\left(
\genfrac{}{}{0pt}{}{\varepsilon_{R}^{MN}\Gamma^{MN}}{0}%
\genfrac{}{}{0pt}{}{0}{R}%
\right)  , $ one can see that the structures $g\left(
\genfrac{}{}{0pt}{}{\text{ }X_{i}\cdot\Gamma}{0}%
\genfrac{}{}{0pt}{}{0}{0}%
\right)  g^{-1}$ and $X_{i}\cdot X_{j}$ are gauge invariant under $\delta_{R}%
$. Note that $g$ transforms from the right side under $\delta_{R}. $ The local
kappa supersymmetry also acts on $g$ from the right as $\delta_{\kappa}g=gK$
with $K=\left(
\genfrac{}{}{0pt}{}{0}{\bar{\kappa}^{i}\Gamma_{M}}%
\genfrac{}{}{0pt}{}{\Gamma_{M}\kappa^{i}}{0}%
\right)  X_{i}^{M},$ and on $\delta_{\kappa}A^{ij}\neq0$ as follows. Under
this $\delta_{\kappa}$ transformation the action in the form of Eq.(\ref{S2T})
gives $\delta_{\kappa}S_{2T}=\int d\tau\left[  -\frac{1}{2}\delta_{\kappa
}A^{ij}X_{i}\cdot X_{j}+\frac{4}{s_{d}}Str\left(  i\delta_{\kappa}\left(
g^{-1}\partial_{\tau}g\right)  L\right)  \right]  ,$ where the second term
takes the form $Str\left(  ig^{-1}\partial_{\tau}g\left(
\genfrac{}{}{0pt}{}{0}{\bar{\sigma}}%
\genfrac{}{}{0pt}{}{\sigma}{0}%
\right)  \right)  $ with $\sigma=\left(  \Gamma_{NK}\Gamma_{M}\kappa
^{i}\right)
X_{i}^{M}X_{j}^{N}X_{k}^{K}\varepsilon^{jk}=-2X_{i}\cdot X_{j}$
$\Gamma_{N}\kappa^{i}X_{k}^{N}\varepsilon^{jk}$. The important thing
is that $Str\left(  i\delta_{\kappa}\left(
g^{-1}\partial_{\tau}g\right)  L\right)  $ is proportional to
$X_{i}\cdot X_{j},$ and therefore it can be canceled by choosing
$\delta_{\kappa}A^{ij}$ in front of the same coefficient $X_{i}\cdot
X_{j},$ so that $\delta_{\kappa}S_{2T}=0.$

The action $S_{2T}\left(  X,P,g\right)  $ can be generalized by increasing the
number of dimensions, but keeping the same $g\in G_{\text{super}}.$ We will
denote the new action as $S_{2T}\left(  \hat{X},\hat{P},g\right)  $. To
describe its content, first we recall that the group element $g\in
G_{\text{super}}$ was chosen by considering the number of dimensions $d$ and a
group $G\supset$SO$\left(  d,2\right)  $ as listed in Table 1. Now we extend
the $d+2 $ dimensions $X_{i}^{M}=\left(  X^{M},P^{M}\right)  $ by adding
$d^{\prime}$ more spacelike dimensions $X_{i}^{I}=\left(  X^{I},P^{J}\right)
,$ $I=1,2,\cdots,d^{\prime}.$ We associate the SO$\left(  d^{\prime}\right)  $
acting on the $d^{\prime}$ dimensions with the $R$-symmetry group, just as the
$d+2$ dimensions are associated with the group $G.$ Namely we choose the
number of dimensions $d^{\prime}$ such that the dimension $s_{d^{\prime}}$ of
the spinor representation of SO$\left(  d^{\prime}\right)  $ coincides with
the fundamental representation of the R-symmetry group. Then, instead of the
$L$ in Eq.(\ref{L}) we define an extended $\hat{L}$
\begin{equation}
\hat{L}=\frac{1}{4i}\left(
\begin{array}
[c]{cc}%
\Gamma^{MN}L_{MN} & 0\\
0 & -\alpha\Gamma^{IJ}L_{IJ}%
\end{array}
\right)  \label{Lhat}%
\end{equation}
where $\alpha=\frac{s_{d}}{s_{d^{\prime}}}$ is the ratio of the dimensions of
the spinor representations for SO$\left(  d+2\right)  $ and SO$\left(
d^{\prime}\right)  .$ Now define $\hat{X}_{1}^{\hat{M}}=\left(  X^{M}%
,X^{I}\right)  $ and $\hat{X}_{2}^{\hat{M}}=\left(
P^{M},P^{I}\right)  $ as the phase space in $d+d^{\prime}+2$
dimensions, and write the same form of action as Eq.(\ref{S2T}) in
the \textit{extended} dimensions $S\left(  \hat {X},\hat{P},g\right)
,$ but use the new form of $\hat{L}$ given above. The coupling
$A^{ij}\hat{X}_{i}\cdot\hat{X}_{j}$ leads to the Sp$\left(
2,R\right)  $ constraints that includes all dimensions at an equal
footing. In the coupling $Str\left(  ig^{-1}\partial_{\tau}gL\right)
$ the first $d+2 $ dimensions couple to the SO$\left(  d,2\right)  $
in $G$ and the last $d^{\prime}$ dimensions couple to SO$\left(
d^{\prime}\right)  $ in the $R$-symmetry group. This extended action
has the following local and global
symmetries%
\begin{equation}
\text{local: Sp}\left(  2,R\right)  \times\left(
\genfrac{}{}{0pt}{}{\text{ SO}\left(  d,2\right)  }{\frac{1}{2}\kappa appa}%
\genfrac{}{}{0pt}{}{\frac{1}{2}\kappa appa}{\text{SO}\left(  d^{\prime
}\right)  }%
\right)  ,\;\text{global: }G_{\text{super}} \label{localglobal2}%
\end{equation}
Note that the global symmetry is the same as before the extension since the
Cartan connection $g^{-1}\partial_{\tau}g$ is still invariant, but the local
symmetry is less as seen by comparing to Eq.(\ref{localglobal1}). In
particular now we have $\frac{1}{2}\kappa appa$ so more fermions in $g$ are
physical. Also there may be more bosons if SO$\left(  d^{\prime}\right)  $ is
smaller than the R-symmetry in $G_{\text{super}}$ as labeled by $N$ in Table
1. So, the extended model is expected to have a different physical set of
states. In this scheme we obtain interesting cases, depending on the gauge
choices, such as superparticles with compactified $d^{\prime}$ dimensions, and
without D-branes for the cases $d=3,4,5,6$. With the right choice of
dimensions and groups the emerging space is quite interesting from the point
of view of $M$-theory.

For example \cite{2tAdSs}\cite{2tsuperstring}\cite{2tstringtwistors} the
superparticle on AdS$_{5}\times$S$^{5}$ with a total of 10 dimensions ($d=4$,
$d^{\prime}=6,$ and $d+d^{\prime}+2=12$) is obtained by taking the supergroup
SU$\left(  2,2|4\right)  ,$ and then specializing to a particular gauge. It
was shown in \cite{2tAdSs} that this approach gives a particle spectrum that
is identical to the Kaluza-Klein towers of $d=10$ supergravity compactified on
AdS$_{5}\times$S$^{5}.$ This was discussed by choosing the spacetime gauge as
shown in the first branch of Fig.2. The same theory can be brought to the form
of a sigma model for the coset SU$\left(  2,2|4\right)  /$SO(4,1)$\times
$SO$\left(  5\right)  $ \cite{tseytlin}, as shown in the third branch of
Fig.2, or can be put to the form of a twistor theory as shown in the second
branch of Fig.2.

Similarly the superparticle on AdS$_{4}\times$S$^{7}$ ($d=3,$ $d^{\prime}=8)$
or AdS$_{7}\times$S$^{4}$ ($d=6,$ $d^{\prime}=5)$ with a total of 11
dimensions, and no D-branes, emerges by taking the supergroup OSp$\left(
8|4\right)  $ \cite{2tsuperstring}\cite{2tstringtwistors}. More details will
be given in a separate publication.

\subsection{Covariant quantization \& representations of G$_{\text{super}}$}

As seen from the form of $J$ in Eq.(\ref{J}), it is gauge invariant under
$\left(
\genfrac{}{}{0pt}{}{\text{ SO}\left(  d,2\right)  }{\text{super}}%
\genfrac{}{}{0pt}{}{\text{super}}{\text{R-symm}}%
\right)  $ as well as Sp$\left(  2,R\right)  $ transformations. Therefore the
$G_{\text{super}}$ charges $J_{A}^{~B}$ are physical observables that classify
the physical states under $G_{\text{super}}$ representations. With this in
mind we study the properties of $J.$ In particular the square of the matrix
$J,$ given by $\left(  J^{2}\right)  _{A}^{~B}=\left(  gLg^{-1}gLg^{-1}%
\right)  _{A}^{~~B}=\left(  gL^{2}g^{-1}\right)  _{A}^{~~B},$ contains
important information about the physical states. At the classical level
$L^{2}=0$ as discussed in section (\ref{L2section}), and therefore $\left(
J^{2}\right)  _{A}^{~B}=0$ at the classical level. At the quantum level we
must be careful not only about the computation of $L^{2}$ as discussed in
section (\ref{L2section}), but also about the order of operators in $gLg^{-1}$
because, unlike section (\ref{L2section}), $g$ cannot be fully eliminated by
the available gauge symmetries. Then $J^{2}$ is not necessarily of the form
$gL^{2}g^{-1}$ except for the simplest case of $G=$SO$\left(  d,2\right)  $.

The details of the quantum discussion will be given in a separate paper, but
suffice it to mention that we obtain the following general form of algebraic
constraints among the generators of $G_{\text{super}}$
\begin{equation}
JJ=\alpha J+\beta1 \label{JJtoJ}%
\end{equation}
where the coefficients $\alpha,\beta$ depend on $G_{\text{super}}.$ This
equation is to be compared to the simpler case of SO$\left(  d,2\right)  $ in
Eq.(\ref{LLtoL}). By taking a super trace we learn that $\beta$ gives the
quadratic Casimir operator. The absence of any quadratic term in $J$ on the
right hand side of Eq.(\ref{JJtoJ}) not proportional to $1$ is very
nontrivial. This does not happen for generic representations of
$G_{\text{super}}.$ But for the representations generated by the from
$J=gLg^{-1}$ it is expected since $J^{2}$ vanished at the classical level.

So the algebraic constraints above must determine the representation of
$G_{\text{super}}$. Indeed, by using Eq.(\ref{JJtoJ}) repeatedly we obtain%
\begin{equation}
J^{n}=\alpha_{n}J+\beta_{n}1,\;C_{n}\sim Str\left(  \left(  2J\right)
^{n}\right)  \sim2^{n}\beta_{n}.
\end{equation}
where all coefficients $\alpha_{n},\beta_{n}$ and the Casimir eigenvalues
$C_{n}$ are determined in terms of $\alpha,\beta$ as follows%
\begin{equation}
\alpha_{n}=\frac{\left(  \alpha+\sqrt{\alpha^{2}+4\beta}\right)
^{n+1}-\left(  \alpha-\sqrt{\alpha^{2}+4\beta}\right)  ^{n+1}}{2^{n+1}%
\sqrt{\alpha^{2}+4\beta}},\;\beta_{n}=\beta\alpha_{n-2}.\;\;
\end{equation}
These properties of the Casimir eigenvalues completely determine the
representation of $G_{\text{super}}$.

This is the representation that classifies the physical states of the theory
$S_{2T}\left(  X,P,g\right)  ,$ or the extended one $S_{2T}\left(  \hat
{X},\hat{P},g\right)  ,$ under the global symmetry $G_{\text{super}}.$ No
matter which gauge we choose to describe the physical content of the theory we
cannot change the group theoretical content of the physical states. In the
particle gauge, in position space these physical states correspond to an
on-shell free field in a field theory. Some cases of interest are listed in
the table below. These were obtained by covariant quantization of the 2T
particles or superparticles in various dimensions without choosing a gauge.
Details of the computation will appear elsewhere
\[%
\begin{tabular}
[c]{|lll|}\hline
{\small particle/superparticle} & \multicolumn{1}{|l}{G$_{\text{super}}$
algebraic constraints} & \multicolumn{1}{|l|}{Field theory}\\\hline
$%
\genfrac{}{}{0pt}{}{\text{SO}\left(  d,2\right)  \text{, any }%
d\;~~~\;\;}{\text{massless, spinless\ }}%
$ & \multicolumn{1}{|l}{$JJ=-\frac{d}{2}J+\frac{1}{4}\left(  1-\frac{d^{2}}%
{4}\right)  $} & \multicolumn{1}{|l|}{{\small Scalar Klein-Gordon}}\\\hline
$%
\genfrac{}{}{0pt}{}{\text{SO}\left(  4,2\right)  \text{, }%
d=4\;\;~~~~\;\;\;\;\;\;\;\;\;}{\text{massless, any helicity~}h\text{\ }}%
$ & \multicolumn{1}{|l}{$JJ=\left(  h-2\right)  J+\frac{3}{4}\left(
h^{2}-1\right)  $} & \multicolumn{1}{|l|}{$%
\genfrac{}{}{0pt}{}{\text{massless field, any spin~~~}}{\text{Dirac, Maxwell,
Einstein,}..}%
$}\\\hline
$\text{OSp}\left(  N|4\right)  ,\text{ }d=3$ & \multicolumn{1}{|l}{$%
\genfrac{}{}{0pt}{}{N=8\;\;JJ=-\frac{3}{2}J+\beta_{8}~~\;}{N=16\;\;JJ=-\frac
{3}{2}J+\beta_{16}}
$} & \multicolumn{1}{|l|}{$%
\genfrac{}{}{0pt}{}{N=8\;\text{Super Yang-Mills}%
}{N=16\;\text{SUGRA~~~~~~~~~~~~~~~}}%
$}\\\hline
$\text{SU}\left(  2,2|N\right)  ,\text{ }d=4$ & \multicolumn{1}{|l}{$%
\genfrac{}{}{0pt}{}{N=4\;\;JJ=-2J+0~~~}{N=8\;\;JJ=-2J+\frac{5}{4}~}%
$} & \multicolumn{1}{|l|}{$%
\genfrac{}{}{0pt}{}{N=4\;\text{Super Yang-Mills}%
}{N=8\;\text{SUGRA~~~~~~~~~~~~~~~~}}%
$}\\\hline
$\text{OSp}\left(  8^{\ast}|N\right)  ,\text{ }d=6$ & \multicolumn{1}{|l}{$%
\genfrac{}{}{0pt}{}{N=4\;\;JJ=-3J+\beta_{4}\;\;~~\;}{N=8\;\;JJ=-3J+\beta
_{8}~\;\;~~}%
$} & \multicolumn{1}{|l|}{$%
\genfrac{}{}{0pt}{}{N=4\;\;\text{self-dual CFT}%
}{N=8~~~~~\text{\ \ ~~~~~~~~~~~~~~~~~~\ }}%
$}\\\hline
$\text{SU}\left(  2,2|4\right)  ,%
\genfrac{}{}{0pt}{}{d=4~}{d^{\prime}=6\text{ \ }}%
$ & \multicolumn{1}{|l}{$JJ=-2J+\frac{l\left(  l+1\right)  }{4}%
,\;l{\small =1,2,3,\cdots}$} & \multicolumn{1}{|l|}{$%
\genfrac{}{}{0pt}{}{\text{type IIB, AdS}_{5}\times\text{S$^{5}$~\ \ \ }%
}{\text{compactified\ SUGRA}}%
$}\\\hline
\multicolumn{3}{|l|}{\ \ \ \ \ {\small Table 2 - Algebraic constraints
}$JJ=\alpha J+\beta1${\small \ satisfied by the generators of }%
$G_{\text{super}}$}\\\hline
\end{tabular}
\]
If we choose other gauges than the particle gauge, we find other holographic
images of the same 2T-theory. The other images are dual to the particle or the
field theory image included in the table above. The various gauges will yield
the same representation of $G_{\text{super}}$ since the Casimir eigenvalues
are gauge invariant and cannot change. The quantum states in various images
can differ from one image to another only by the set of operators that are
simultaneously diagonal on the physical state (usually including the
Hamiltonian for that image) beyond the Casimir operators. Since a gauge
transformation is a duality transformation from one image to another, this
duality transformation is a unitary transformation within the unitary
representation of $G_{\text{super}}$ fixed above by $\alpha,\beta.$

\subsection{Twistor gauge: supertwistors dual to super phase space
\label{stwgaugeSec}}

There are different ways of choosing gauges to express the theory
given by $S_{2T}\left(  X,P,g\right)  ,$ or the extended one
$S_{2T}(  \hat {X},\hat{P},g)  $, in terms of the physical sector.
One extreme in gauge space is to eliminate all of the
SO(d,2)$\times$R-symm subgroup of $g$ completely, while another
extreme is to eliminate $\left( X,P\right) $ completely. When most
of $g$ is eliminated we obtain the phase space description, and when
$\left( X,P\right) $ is eliminated we obtain the twistor
description.

To obtain the twistor description for the action $S_{2T}\left(  X,P,g\right)
$ we eliminate $\left(  X^{M},P^{M}\right)  $ completely and keep only $g$ as
discussed in \cite{2ttwistor}. This is done by using the Sp$\left(
2,R\right)  $ and the SO$\left(  d,2\right)  $ local symmetries to completely
fix $X^{M},P^{M}$ to the convenient form $X^{+^{\prime}}=1$ and $P^{+}=1,$
while all other components vanish%
\begin{equation}
X^{M}=(\overset{+^{\prime}}{{1}},\overset{-^{\prime}}{{0}},\overset{+}{{0}%
},\overset{-}{{0}},\overset{i}{{0}}),\;P^{M}=(\overset{+^{\prime}}{{0}%
},\overset{-^{\prime}}{{0}},\overset{+}{{1}},\overset{-}{{0}},\overset{i}{{0}%
}),\;i=1,\cdots,\left(  d-2\right)  . \label{twistgauge1}%
\end{equation}
These $X^{M},P^{M}$ already satisfy the constraints $X^{2}=P^{2}=X\cdot P=0$.
In this gauge the only non-vanishing component of $L^{MN}$ is $L^{+^{\prime}%
+}=1$, so that
\begin{equation}
L_{fixed}=\frac{-2}{4i}\left(
\genfrac{}{}{0pt}{}{\text{ }\Gamma^{-^{\prime}-}}{0}%
\genfrac{}{}{0pt}{}{0}{0}%
\right)  L^{+^{\prime}+}=\frac{i}{2}\left(
\genfrac{}{}{0pt}{}{\text{ }\Gamma^{-^{\prime}-}}{0}%
\genfrac{}{}{0pt}{}{0}{0}%
\right)  \equiv\Gamma. \label{Lfixed}%
\end{equation}
Hence the physical content of the theory is now described only in terms of $g
$ and the fixed matrix $\Gamma$ embedded in the Lie algebra of SO$\left(
d,2\right)  .$

The matrix $\Gamma$ has very few non-zero entries as seen by choosing a
convenient form of gamma matrices\footnote{An explicit form of SO$\left(
d,2\right)  $ gamma matrices that we find convenient in even dimensions, is
given by $\Gamma^{0}=-1\times1$,\ $\Gamma^{i}=\sigma_{3}\times\gamma^{i}%
$,\ $\Gamma^{\pm^{\prime}}=-i\sqrt{2}\sigma^{\pm}\times1$ (note $\Gamma
^{0^{\prime}}=-i\sigma_{1}\times1$ and $\Gamma^{1^{\prime}}=\sigma_{2}\times
1$), \ where $\gamma^{i}$ are the SO$\left(  d-1\right)  $ gamma matrices. The
$\bar{\Gamma}^{M}$ are the same as the $\Gamma^{M}$ for $M=\pm^{\prime},i,$
but for $M=0$ we have $\bar{\Gamma}^{0}=-\Gamma^{0}=1\times1.$ From these we
construct the traceless $\Gamma^{+^{\prime}-^{\prime}}=\left(
\genfrac{}{}{0pt}{}{-1}{0}%
\genfrac{}{}{0pt}{}{0}{1}%
\right)  $,\ $\Gamma^{+^{\prime}\mu}=i\sqrt{2}\left(
\genfrac{}{}{0pt}{}{0}{0}%
\genfrac{}{}{0pt}{}{\bar{\gamma}^{\mu}}{0}%
\right)  $,\ $\Gamma^{-^{\prime}\mu}=-i\sqrt{2}\left(
\genfrac{}{}{0pt}{}{0}{\gamma^{\mu}}%
\genfrac{}{}{0pt}{}{0}{0}%
\right)  $,\ $\Gamma^{\mu\nu}=\left(
\genfrac{}{}{0pt}{}{\bar{\gamma}^{\mu\nu}}{0}%
\genfrac{}{}{0pt}{}{0}{\gamma^{\mu\nu}}%
\right)  $, with $\gamma^{\mu}=\left(  1,\gamma^{i}\right)  $ and $\bar
{\gamma}^{\mu}=\left(  -1,\gamma^{i}\right)  .$ Then $\frac{1}{2}\Gamma
_{MN}J^{MN}=-\Gamma^{+^{\prime}-^{\prime}}J^{+^{\prime}-^{\prime}}$+~
$\frac{1}{2}J_{\mu\nu}\Gamma^{\mu\nu}-$ $\Gamma_{~\mu}^{+^{\prime}%
}J^{-^{\prime}\mu}-$ $\Gamma_{~\mu}^{-^{\prime}}J^{+^{\prime}\mu}$ takes the
matrix form given in Eq.(\ref{GL}). We can further write $\gamma^{1}=\tau
^{1}\times1,$ $\gamma^{2}=\tau^{2}\times1$ and $\gamma^{r}=\tau^{3}\times
\rho^{r}$ for $\rho^{r}$ the gamma matrices for SO$\left(  d-3\right)  $.
These gamma matrices are consistent with the metric $C=\sigma_{1}\times1\times
c$ of Eq.(\ref{C}), and footnote (\ref{ginverse}), provided $c^{-1}\left(
\rho^{r}\right)  ^{\dagger}c=\rho^{r}.$ It is possible to choose hermitian
$\rho^{r}$ with $c=1$ for SO$\left(  d-3\right)  .$ If one works in a basis
with $c\neq1,$ then hermitian conjugation of of SO$\left(  d-3\right)  $
spinors (which occur e.g. in $\bar{\lambda}$ of Eq.(\ref{zbar})) must be
supplemented by multiplying with $c,$ as in $\bar{\lambda}\equiv
\lambda^{\dagger}\left(  1\times c\right)  .$ \label{gamms}} for SO$\left(
d,2\right)  $. Then, up to similarity transformations, $\Gamma$ can be brought
to the form\footnote{The gamma matrices $\Gamma^{M}$ of footnote (\ref{gamms})
can be redefined differently for the left or right sides of $g$ up to
similarity transformations. Thus, for the right side of $g$ we apply a
similarity transformation so that $\gamma^{1}=\tau^{3}\times1$, etc., to
obtain $\gamma^{-}=\left(  \gamma^{0}-\gamma^{1}\right)  /\sqrt{2}$ in the
form given in Eq.(\ref{g--}).}
\begin{equation}
\Gamma=\frac{1}{\sqrt{2}}\left(
\begin{array}
[c]{cc}%
\genfrac{}{}{0pt}{}{0}{\gamma^{-}}%
\genfrac{}{}{0pt}{}{0}{0}%
& 0\\
0 & 0
\end{array}
\right)  =\left(
\begin{array}
[c]{cc}%
\genfrac{}{}{0pt}{}{\genfrac{}{}{0pt}{}{0}{0}\genfrac{}{}{0pt}{}{0}{0}%
}{\genfrac{}{}{0pt}{}{0}{0}\genfrac{}{}{0pt}{}{0}{1}}%
\genfrac{}{}{0pt}{}{\genfrac{}{}{0pt}{}{0}{0}\genfrac{}{}{0pt}{}{0}{0}%
}{\genfrac{}{}{0pt}{}{0}{0}\genfrac{}{}{0pt}{}{0}{0}}%
& 0\\
0 & 0
\end{array}
\right)  .\label{g--}%
\end{equation}
The identity matrix $1,$ and the small $0$'s in the last expression are
$\frac{s_{d}}{4}\times\frac{s_{d}}{4}$ square block matrices embedded in the
$s_{d}\times s_{d}$ spinor representation of SO$\left(  d,2\right)  .$ Then
the gauge invariant 2T action in Eq.(\ref{S2T}), and the gauge invariant
SO$\left(  d,2\right)  _{L}$ charges in Eq.(\ref{J}), take the twistor form
similar to Eq.(\ref{action})
\begin{align}
S_{2T}\left(  X,P,g\right)   &  =\frac{4}{s_{d}}\int d\tau~Str\left(
i\partial_{\tau}g\Gamma g^{-1}\right)  =\frac{4}{s_{d}}\int d\tau~i\bar{Z}%
_{a}^{~A}\partial_{\tau}Z_{A}^{~a}\equiv S_{twistor},\label{Stw}\\
J_{A}^{~B} &  =\left(  g\Gamma g^{-1}\right)  _{A}^{~B}=\left(  Z_{A}^{~a}%
\bar{Z}_{a}^{~B}-\frac{1}{s_{d}-s_{d^{\prime}}}Str\left(  Z\bar{Z}\right)
\delta_{A}^{~B}\right)  ,\;\label{Jtw}%
\end{align}
The $Z_{A}^{~a},\bar{Z}_{a}^{~B},$ as constructed from the group element $g,$
are supertwistors that already obey constraints as we explain below, so
$\frac{4}{s_{d}}\int d\tau~i\bar{Z}_{a}^{~A}\partial_{\tau}Z_{A}^{~a}$ is the
full supertwistor action. Due to the form of $\Gamma$ it is useful to think of
$g$ as written in the form of $\frac{s_{d}}{4}\times\frac{s_{d}}{4}$ square
blocks. Then $Z_{A}^{~a}$ with $A=$ fundamental of $G_{\text{super}}$ and
$a=1,2,\cdots,\frac{s_{d}}{4}$ emerges as the rectangular supermatrix that
corresponds to the fourth block of columns of the matrix $g,$ and similarly
$\bar{Z}_{a}^{~A}$ corresponds to the second block of rows of $g^{-1}.$ Since
$g^{-1}=\left(
\begin{array}
[c]{cc}%
C^{-1} & 0\\
0 & 1
\end{array}
\right)  g^{\dagger}\left(
\begin{array}
[c]{cc}%
C & 0\\
0 & 1
\end{array}
\right)  ,$ we find that $\bar{Z}=c^{-1}Z^{\dagger}\left(
\begin{array}
[c]{cc}%
C & 0\\
0 & 1
\end{array}
\right)  ,$ where $C=\sigma_{1}\times1\times c$ is given in footnote
(\ref{gamms}). Furthermore, as part of $g,g^{-1},$ the $Z_{A}^{~a},\bar{Z}%
_{a}^{~B}$ must satisfy the constraint $\bar{Z}_{a}^{~A}Z_{A}^{~b}=0$ since
the product $\bar{Z}_{a}^{~A}Z_{A}^{~b}$ contributes to an off-diagonal block
of the matrix $1$ in $g^{-1}g=1,\;$
\begin{equation}
g^{-1}g=1\;\rightarrow\bar{Z}_{a}^{~A}Z_{A}^{~b}=0.\label{ZZconstraint}%
\end{equation}
A constraint such as this one must be viewed as the generator of a gauge
symmetry that operates on the $a$ index (the columns) of the supertwistor
$Z_{A}^{~a}.$ For the purely bosonic version of this process see
\cite{twistorBP2} where twistors in any dimension are obtained.

In $d=4,6$ the supertwistors that emerge from this approach coincide with
supertwistors previously known in the literature if we work in the massless
particle gauge \cite{2ttwistor}\cite{2tsuperstring}\cite{2tstringtwistors}.
However, if we work in one of the other gauge choices that lead to the
holographic images depicted in Figs.1,2, then we obtain new results for the
twistor description for those cases. Some applications of the $d=4,6$ twistors
will be given in the next section.

Similarly, to obtain the twistor description for the extended action
$S_{2T}\left(  \hat{X},\hat{P},g\right)  $ we eliminate $\left(  \hat{X}%
^{\hat{M}},\hat{P}^{\hat{M}}\right)  $ completely and keep only $g$ as
discussed in \cite{2tsuperstring}\cite{2tstringtwistors}. Thus, we first use
the local SO$\left(  d,2\right)  \times$SO$\left(  d^{\prime}\right)  \subset
$G$_{\text{super}}$ to rotate the $d+d^{\prime}+2$ components to the form
\begin{align}
\hat{M} &  =\left(  \,\,0^{\prime}~~0~~~1~\cdots~d~~,~I=1\,~2~~3~~\cdots
~d^{\prime}\right)  \nonumber\\
\hat{X}^{\hat{M}}\left(  \tau\right)   &  =\left(  \,\,1~~~0~~~0~~\cdots
~0~~,~~~~1\,~~~0~~~~0~~\cdots~0\right)  \\
\hat{P}^{\hat{M}}\left(  \tau\right)   &  =\left(  \,\,0~~~1~~~0~~\cdots
~0~~,~~~~0\,~~~1~~~~0~~\cdots~0\right)
\end{align}
These solve also the Sp(2,R) constraints. In this gauge the extended matrix
$\hat{L}$ simplifies to%
\begin{equation}
\hat{L}_{fixed}\sim\left(
\begin{array}
[c]{cc}%
i\Gamma_{0^{\prime}0} & 0\\
0 & -i\alpha\Gamma_{12}%
\end{array}
\right)  =\left(
\begin{array}
[c]{cccc}%
1_{s_{d/2}} & 0 & 0 & 0\\
0 & -1_{s_{d/2}} & 0 & 0\\
0 & 0 & -\alpha1_{s_{d^{\prime}/2}} & 0\\
0 & 0 & 0 & \alpha1_{s_{d^{\prime}/2}}%
\end{array}
\right)  \equiv\hat{\Gamma}%
\end{equation}
In this gauge the action and the $G_{\text{super}}$ symmetry current are
expressed only in terms of the group element%
\begin{align}
S_{2T}\left(  \hat{X},\hat{P},g\right)   &  \sim\int Str\left(  g^{-1}%
\hat{\Gamma}i\partial g\right)  ,\;\;J_{A}^{~B}=\left(  g^{-1}\hat{\Gamma
}g\right)  _{A}^{~B}\label{Jhattwist}\\
g &  \in\text{{\ }}{G}_{\text{super}}\text{{~/~H}}_{\hat{\Gamma}}%
\end{align}
Due to the form of $\hat{\Gamma}$ there are gauge symmetries {H}$_{\hat
{\Gamma}}$ that correspond to all generators of {\ }${G}_{\text{super}}$ that
commute with $\hat{\Gamma}.$ The gauge symmetries remove degrees of freedom so
that the physical degrees of freedom that remain in $g$ corresponds to the
coset ${G}_{\text{super}}${~/~H}$_{\hat{\Gamma}}${. As shown in
\cite{2tsuperstring}\cite{2tstringtwistors} in the case of }${G}%
_{\text{super}}=${PSU}${(2|2)}$ the coset is $\frac{{PSU(2,2|4)~}}%
{{PSU(2|2)}\times{PSU(2|2)}}${\ as seen for }$\hat{\Gamma}=$diag$\left(
1,1,1,1,-1,-1,-1,-1\right)  $ after a rearrangement of rows and columns.
{These coset degrees of freedom are equivalent to the superparticle moving on
AdS}$_{5}\times$S$^{5}$ space.

Furthermore, by an appropriate parametrization of $g,$ including the gauge
degrees of freedom, this action can be written in twistor form. The twistors
in this case is a gauged super grassmanian $Z_{A}^{~a}$ described as follows%
\begin{equation}%
\begin{tabular}
[c]{|l|}\hline
$\overset{{A=1,\cdots,8}}{Z_{A}^{~a}~=}(%
\genfrac{}{}{0pt}{}{\overset{a=1,2}{{bose}}}{{fermi}}%
\genfrac{}{}{0pt}{}{\overset{a=3,4}{{fermi}}}{{bose}}%
)\;\;\;%
\genfrac{}{}{0pt}{}{\text{{8x4 rectangular matrix\ \ \ \ \ \ \ \ \ \ \ }%
}}{\text{{4 fundamental reps of PSU(2,2$|$4)}}}%
\genfrac{}{}{0pt}{}{\text{{\ }}}{^{\text{{\ }}}}%
$\\\hline
$Z_{A}^{~a}~=\text{{\ (8,4)} {\ of }}{PSU(2,2|4)}_{\text{\emph{global}}}%
\times{[PSU(2|2)\times U(1)]}_{\text{\emph{local}}}$\\\hline
$L=\bar{Z}_{a}^{~A}\left(  \left(  \partial+V\right)  Z\right)  _{A}%
^{~a},~~_{\text{{\ }}}V_{a}^{~b}={PSU(2|2)\times U(1)}$ gauge field\\\hline
$J_{A}^{~B}=\left(  Z\bar{Z}-l\right)  _{A\text{{\ }}}^{~B}$ , $l\equiv
\frac{1}{8}Str\left[  \left(
\genfrac{}{}{0pt}{}{1}{0}%
\genfrac{}{}{0pt}{}{0}{-1}%
\right)  \bar{Z}Z\right]  \;\;\;\;%
\genfrac{}{}{0pt}{}{~~~~~~~~\text{global symmetry }{PSU(2,2|4)}%
}{\text{classifies physical states}}%
$\\\hline
$\text{U}\left(  1\right)  :\;\;\;\;\;\Delta\equiv Str\left(  \bar{Z}Z\right)
=Str\left(  Z\bar{Z}\right)  =0~\;\;\;\;\;\;$\ $%
\genfrac{}{}{0pt}{}{~~~~\text{vanish on}~\text{{gauge invariants}}}{\text{{in
twistor space~~~~~~~~~~~~}}}%
$\\\hline
PSU$\left(  2|2\right)  :~G_{a}^{~b}\equiv\bar{Z}_{a}^{~A}Z_{A}^{~b}%
-2l\delta_{a}^{~b}-\frac{\Delta}{4}\left(
\genfrac{}{}{0pt}{}{1}{0}%
\genfrac{}{}{0pt}{}{0}{-1}%
\right)  _{a}^{~b}=0~%
\genfrac{}{}{0pt}{}{~~~~\text{vanish on}~\text{{gauge invariants}}}{\text{{in
twistor space~~~~~~~~~~~~}}}%
$\\\hline
\end{tabular}
\ \ \ \label{Zadss10}%
\end{equation}
The ${PSU(2|2)\times U(1)]}_{\text{\emph{local}}}$ gauge invariant physical
space described by this twistor is the full Kaluza-Klein spectrum of of
type-IIB d=10 supergravity compactified on {\ AdS}$_{5}\times$S$^{5}$, which
is classified by ${PSU(2,2|4)}_{\text{\emph{global}}}$ representations
labeled by the eigenvalues of the operator $l=0,1,2,\cdots.$ This result was
already obtained in \cite{2tAdSs} in a different gauge of the same 2T-physics
action $S_{2T}\left(  \hat{X},\hat{P},g\right)  $.

In a similar way, when ${G}_{\text{super}}=$OSp$\left(  8|4\right)  ,$ the
coset is $\frac{\text{OSp}\left(  8|4\right)  }{\text{Osp}\left(  4|2\right)
\times\text{Osp}\left(  4|2\right)  \times\text{U}\left(  1\right)
\times\text{U}\left(  1\right)  }$ for $\Gamma=$diag$\left(
1,1,1,1,-1,-1,-1,-1,-2,-2,2,2\right)  $. The twistor equivalent version is a
gauged super Grassmanian $Z_{A}^{~a}$ described as follows%
\begin{gather}
Z_{A}^{~a}=\overset{4_{B}\;\;\;\;\;\;\;\;\;\;2_{F}}{%
\genfrac{}{}{0pt}{}{8_{B}}{4_{F}}%
\left(
\genfrac{}{}{0pt}{}{\left(  8\times4\right)  _{B}}{\left(  4\times4\right)
_{F}}%
\genfrac{}{}{0pt}{}{\left(  8\times2\right)  _{F}}{\left(  4\times2\right)
_{B}}%
\right)  ,\;}\text{pseudo-real }Z_{A}^{~a}\label{Zadss11}\\
\text{OSp}\left(  8|4\right)  \text{ global acting on }A,\;\text{OSp}\left(
4|2\right)  \times R\;\text{ local acting on }a\nonumber\\
L=iStr\left(  \bar{Z}DZ\right)  ,\;DZ=\partial_{\tau}Z+AZ,\;A=\text{OSp}%
\left(  4|2\right)  \text{ gauge field}%
\end{gather}
The physical space for this twistor corresponds to 11-dimensional supergravity
compactified on AdS$_{7}\times$S$^{4}$ or AdS$_{4}\times$S$^{7}.$

We have used 2T-physics as a tool to obtain supertwistors that describe
various systems in higher dimensions. Some of the properties of these more
exotic twistors have been outlined in \cite{2tsuperstring}%
\cite{2tstringtwistors}, and more will be discussed elsewhere.

\section{Supertwistors and some field theory spectra in d=4,6}

\subsection{Supertwistors for d=4, N=4 Super Yang-Mills}

Consider the twistor obtained from the 2T-physics approach using
$S_{2T}\left(  X,P,g\right)  $ in $d=4$ and
$G_{\text{super}}=$PSU$\left( 2,2|4\right)  $ as given in Table 1.
The same theory in a different gauge gives the $d=4$ superparticle
with $N=4$ supersymmetries, as described in Eq.(\ref{sparticle}).
The quantum spectra of both descriptions, corresponding to the
physical states, must coincide. Let's see how this is obtained
explicitly.

To begin the superparticle in Eq.(\ref{sparticle}) has $4x$,$4p$ and
16$\theta$ real degrees of freedom in super phase space. In the lightcone
gauge we remove $1x$ and $1p,$ due to $\tau$ reparametrization and the
corresponding $p^{2}=0$ constraint. We also remove $8$ fermionic degrees of
freedom due to kappa supersymmetry. We are left over with $3x,3p,8\theta$
physical degrees of freedom. With these we construct the physical quantum
states as an arbitrary linear combination of the basis states in momentum
space $|\vec{p},\alpha\rangle,$ where $\alpha$ is the basis for the Clifford
algebra satisfied by the $8\theta.$ This basis has 8 bosonic states and 8
fermionic states labeled by $\alpha$. Viewed as probability amplitudes in
position space $\langle x,\alpha|\psi\rangle$ these are equivalent to fields
$\psi\left(  x\right)  _{8_{B}+8_{F}}$ which correspond to the independent
\textit{solutions} of all the constraints. One finds that these are the same
as the 8 bose and 8 fermi fields of the Super Yang- Mills (SYM) theory which
are the solutions of the linearized equations of motion in the lightcone
gauge. They consist of two helicities of the gauge field $A_{\pm1}\left(
x\right)  ,$ two helicities for the gauginos $\psi_{+\frac{1}{2}}^{a}\left(
x\right)  ,$ $\bar{\psi}_{-\frac{1}{2},a}\left(  x\right)  $ in the
$\mathbf{4,\bar{4}}$ of SU$\left(  4\right)  ,$ and six scalars $\phi^{\lbrack
ab]}\left(  x\right)  $ in the $\mathbf{6}$ of SU$\left(  4\right)  .$

Now we count the physical degrees of freedom for the twistors. As seen from
Eq.(\ref{Stw}), for $d=4$ we have one complex twistor $Z_{A}$ in the
fundamental representation of PSU$\left(  2,2|4\right)  ,$ with a Lagrangian
and a conserved current given by
\begin{equation}
\underset{Z_{A}\text{{\ \ is in fundamental representation of PSU(2,2%
$\vert$%
N)~}}\leftrightarrow\text{CP}^{3|N}}{L=i\bar{Z}^{A}\partial_{\tau}%
Z_{A},\;J_{A}^{~B}=Z_{A}\bar{Z}^{B},\;\text{and\ }\bar{Z}^{A}Z_{A}=0}
\label{4Dtwist}%
\end{equation}
This is recognized as the particle version of the $d=4$, $N=4,$ twistor
superstring \cite{witten}-\cite{2tstringtwistors}. To start $Z_{A}$ has $4$
complex bosons and 4 complex fermions, i.e. $8_{B}+8_{F}$ real degrees of
freedom. However, there is one constraint $\bar{Z}^{A}Z_{A}=0$ and a
corresponding U$\left(  1\right)  $ gauge symmetry, which remove 2 bosonic
degrees of freedom. The result is $6_{B}+8_{F}$ physical degrees of freedom
which is equivalent to CP$^{3|4}$, and the same number as $3x,3p,8\theta$ for
the superparticle, as expected.

To construct the spectrum in twistor space we could resort to well known
twistor techniques by working with fields $\phi\left(  Z\right)  $ that are
holomorphic in $Z_{A}$ on which $\bar{Z}^{A}$ acts as a derivative $\bar
{Z}^{A}\phi\left(  Z\right)  =-\partial\phi\left(  Z\right)  /\partial Z_{A},$
as dictated by the canonical structure that follows from the Lagrangian
(\ref{4Dtwist}). Imposing the constraint amounts to requiring $\phi\left(
Z\right)  $ to be homogeneous of degree $0$, namely
\begin{equation}
Z_{A}\frac{\partial\phi\left(  Z\right)  }{\partial Z_{A}}=0,\;Z_{A}%
=\text{PSU}\left(  2,2|4\right)  \text{ supertwistor.}%
\end{equation}
Quantum ordering does not change the homogeneity degree because there are an
equal number of bosons and fermions in the case of $N=4.$ We write
$Z_{A}=\binom{z_{i}}{\xi_{a}}$ with $z_{i}=\left(  \mu,\lambda\right)  $ the
$4$ of SU$\left(  2,2\right)  \subset$PSU$\left(  2,2|4\right)  $ and $\xi
_{a}$ the $4$ of SU$\left(  4\right)  \subset$PSU$\left(  2,2|4\right)  .$
Then the superfield $\phi\left(  Z\right)  $ can be expanded in powers of the
fermions $\xi_{a}$%
\begin{equation}
\phi\left(  Z\right)  =\sum_{n=0}^{4}\left(  \xi_{a_{1}}\cdots\xi_{a_{n}%
}\right)  \phi^{a_{1}\cdots a_{n}}\left(  z\right)  .
\end{equation}
The equation $0=Z_{A}\frac{\partial\phi\left(  Z\right)  }{\partial Z_{A}%
}=z_{i}\frac{\partial\phi\left(  z,\xi\right)  }{\partial z_{i}}+\xi_{a}%
\frac{\partial\phi\left(  z,\xi\right)  }{\partial\xi_{a}}=0$ becomes a
homogeneity condition for the coefficients $\phi^{a_{1}\cdots a_{n}}\left(
z\right)  $
\begin{equation}
z_{i}\frac{\partial\phi^{a_{1}\cdots a_{n}}\left(  z\right)  }{\partial z_{i}%
}=-n\phi^{a_{1}\cdots a_{n}}\left(  z\right)
\end{equation}
Comparing to Eq.(\ref{htwistor}) we see that the helicity of the wavefunction
$\phi^{a_{1}\cdots a_{n}}\left(  z\right)  $ is $h_{n}=\frac{n}{2}-1.$ So we
will label the wavefunction by its helicity as well as its SU$\left(
4\right)  $ labels, by including a subscript $\frac{n}{2}-1$ that corresponds
to the helicity. More explicitly, the wavefunction $\phi\left(  Z\right)  $
takes the form
\begin{align}
\phi\left(  Z\right)   &  =A_{-1}\left(  z\right)  +\xi_{a}\psi_{-1/2}%
^{a}\left(  z\right)  +\xi_{a}\xi_{b}\phi_{0}^{ab}\left(  z\right) \\
&  +\frac{\varepsilon^{abcd}\xi_{a}\xi_{b}\xi_{c}}{3!}\psi_{+1/2,d}\left(
z\right)  +\frac{\varepsilon^{abcd}\xi_{a}\xi_{b}\xi_{c}\xi_{d}}{4!}%
~A_{+1}\left(  z\right)  ,
\end{align}
where we gave suggestive names to the coefficients $\phi^{a_{1}\cdots a_{n}%
}\left(  z\right)  $%
\begin{equation}
\phi^{a_{1}\cdots a_{n}}\left(  z\right)  :\left(  A_{-1},\psi_{-1/2}^{a}%
,\phi_{0}^{ab},\psi_{+1/2,d},A_{+1}\right)  .
\end{equation}
These are precisely the helicity fields, including SU$\left(  4\right)  $
representation content, that correspond to the vector supermultiplet in
$N=4,~d=4$ SYM theory. They each are homogeneous of degree $-2h-2$ where $h$
corresponds to the helicity indicated by the the subscript. For example,
$A_{-1}\left(  z\right)  $ is homogeneous of degree 0, while $A_{+1}\left(
z\right)  $ is homogeneous of degree $-4$, etc. Thus the $\phi\left(
Z\right)  $ is the degree zero wavefunction $\phi\left(  Z\right)  $ described
in \cite{berkWit}. The entire wavefunction can be expanded in terms of
momentum eigenstates as in Eq.(\ref{psipi}) using the results for $\langle
z|k\rangle$ listed in Eq.(\ref{Zk}).

The superfield $\phi\left(  Z\right)  $ is a representation basis of
PSU$\left(  2,2|4\right)  $ which is an evident global symmetry of the twistor
action (\ref{4Dtwist}). The symmetry current $J_{A}^{~B}=Z_{A}\bar{Z}^{B}$
acts as $J_{A}^{~B}\phi\left(  Z\right)  =-Z_{A}\frac{\partial\phi\left(
Z\right)  }{\partial Z_{B}},$ and this induces the symmetry transformations on
the individual fields. This is the hidden superconformal symmetry of the
$d=4,$ $N=4$ SYM field theory in the twistor version. Recall that the twistor
form of $J_{A}^{~B}$ followed by gauge fixing the original gauge invariant
form given in Eq.(\ref{J}), so the PSU$\left(  2,2|4\right)  $ superconformal
symmetry of SYM theory is understood as the global symmetry of the underlying
$4+2$ dimensional superparticle.

Recall that in \cite{berkWit} there are also twistor wavefunctions $f\left(
Z\right)  ,g\left(  Z\right)  $ that describe the spectrum of conformal
gravity; those can arise also in our twistor formalism, but for a different
superparticle model that gives a different degree $c\neq0$ in the PSU$\left(
2,2|4\right)  $ supertwistor homogeneity equation $Z_{A}\frac{\partial\phi
_{k}\left(  Z\right)  }{\partial Z_{A}}=c\partial\phi_{k}\left(  Z\right)  .$
Since only the value of $c=0$ is permitted in the $N=4,~d=4$ superparticle
model, only SYM states $\phi\left(  Z\right)  $ are present. Of course, this
is no surprise in the 2T setting. We have simply compared two gauges, and we
must agree.

\subsubsection{Oscillators, supertwistors, and unitarity of d=4,N=4 spectrum}

It is also worth analyzing the quantum system in terms of oscillators related
to twistors and understand the unitarity of the physical space. We emphasize
that $\bar{Z}^{A}$ is obtained from $Z_{A}$ by Hermitian conjugation and
multiplying by the PSU$\left(  2,2|4\right)  $ metric as given following
Eq.(\ref{Jtw}). To see the oscillator formalism clearly we work in a basis of
SU$\left(  2,2|4\right)  $ in which the group metric is diagonal of the form
$diag\left(  1_{2},-1_{2},1_{4}\right)  .$ The block $diag\left(  1_{2}%
,-1_{2}\right)  =\sigma_{3}\times1$ part is the SU$\left(  2,2\right)  $
metric in the SU$\left(  2\right)  \times$SU$\left(  2\right)  $ basis$,$ to
be contrasted with the SL$\left(  2,C\right)  $ basis in which the metric
$C=\sigma_{1}\times1$ is off-diagonal as in footnote (\ref{gamms}). The two
bases are simply related by a linear transformation that diagonalizes the
SU$\left(  2,2\right)  $ metric $C=\sigma_{1}\times1\rightarrow\sigma
_{3}\times1$. In this diagonal basis we work with compact SU$\left(  2\right)
\times$SU$\left(  2\right)  $ oscillators $z=\left(
\genfrac{}{}{0pt}{}{a_{i}}{\bar{b}^{I}}%
\right)  $ and $\bar{z}=\left(  \bar{a}^{j},-b_{J}\right)  =z^{\dagger}C,$
which are combinations of the SL$\left(  2,C\right)  $ doublet twistor
components $\left(
\genfrac{}{}{0pt}{}{\mu^{\dot{\alpha}}}{\lambda_{\alpha}}%
\right)  ,\left(  \bar{\lambda}_{\dot{\alpha}},\mu^{\alpha}\right)  $ we
discussed before. A bar over the symbol means Hermitian conjugation. In terms
of these, the twistor Lagrangian and the current $J$ take the form%
\begin{align}
L  &  =i\bar{Z}^{A}\partial_{\tau}Z_{A}=i\bar{a}^{i}\partial_{\tau}%
a_{i}-ib_{I}\partial_{\tau}\bar{b}^{I}+i\bar{\xi}^{r}\partial_{\tau}\xi_{r},\;%
\genfrac{}{}{0pt}{}{\overset{\text{SU}\left(  2\right)  }{i=1,2}%
,\;\overset{\text{SU}\left(  2\right)  }{I=1,2}\;}{r=1,\cdots,4\;\text{SU}%
\left(  4\right)  }%
\label{oscillators}\\
Z_{A}  &  =\left(
\begin{array}
[c]{c}%
a_{i}\\
\bar{b}^{I}\\
\xi_{r}%
\end{array}
\right)  ,\;\bar{Z}^{A}=\left(  \bar{a}^{j},-b_{J},\bar{\xi}^{s}\right)
,\;\;\\
J_{A}^{~B}  &  =Z_{A}\bar{Z}^{B}=\left(
\begin{array}
[c]{ccc}%
a_{i}\bar{a}^{j} & -a_{i}b_{J} & a_{i}\bar{\xi}^{s}\\
\bar{b}^{I}\bar{a}^{j} & -\bar{b}^{I}b_{J} & \bar{b}^{I}\bar{\xi}^{s}\\
\xi_{r}\bar{a}^{j} & -\xi_{r}b_{J} & \xi_{r}\bar{\xi}^{s}%
\end{array}
\right)  \label{oscillatorJ}%
\end{align}
It is significant to note that, after taking care of the metric in $\bar{Z}$
as above, the usual canonical rules applied to this Lagrangian identifies the
oscillators as being all \textit{positive norm} oscillators $\left[
a_{i},\bar{a}^{j}\right]  =\delta_{i}^{j},$ $\left[  b_{I},\bar{b}^{J}\right]
=\delta_{I}^{J}$ and $\left\{  \xi_{r},\bar{\xi}^{s}\right\}  =\delta_{r}^{s}$
. Therefore all Fock states have positive norm. In the Fock space of these
oscillators we must identify the physical states as only those that satisfy
the constraints
\[
0=\bar{Z}^{A}Z_{A}=\bar{a}^{i}a_{i}-b_{I}\bar{b}^{I}+\bar{\xi}^{r}\xi_{r}%
=\bar{a}^{i}a_{i}-\left(  \bar{b}^{I}b_{I}+2\right)  +\bar{\xi}^{r}\xi_{r}%
\]
This physical state condition is written in terms of the number operators for
the oscillators $N_{a},N_{\xi},N_{b}$ as
\begin{equation}
\text{physical states:\ }\Delta\equiv N_{a}+N_{\xi}-N_{b}=2
\end{equation}
This setup is precisely the Bars-Gunaydin oscillator formalism for
unitary representations of noncompact groups \cite{barsgunaydin} for
a single \textquotedblleft color\textquotedblright, supplemented
with the constraint $\Delta=2$ as discussed in \cite{2tAdSs}. All
Fock space states that satisfy $\Delta=2$ are easily classified
under the compact subgroup SU$\left( 2\right)  \times$SU$\left(
2\right)  \times$SU$\left(  4\right)  $ of PSU$\left(  2,2|4\right)
.$ They are organized through the unitary infinite dimensional
representations of the subgroup SU$\left(  2,2\right)  $ by
identifying the so called \textquotedblleft
lowest\textquotedblright\ states that are annihilated by the double
annihilation generators $a_{i}b_{J}$ which is part of $J_{A}^{B}$ in
the conformal subgroup SU$\left(  2,2\right)  $ given in
Eq.(\ref{oscillatorJ}). The list of the $\Delta=2$ lowest states is
easily identified and classified under SU$\left(  2\right)
\times$SU$\left( 2\right)  \times$SU$\left(  4\right)
\subset$PSU$\left(  2,2|4\right)  $ as
\begin{equation}
\Delta=2:\left(  \underset{\left(  1,0,1\right)  }{\overset{A_{-1}}{{\bar{a}%
}^{i}{\bar{a}}^{j}}}{,~}\underset{(\frac{1}{2},0,4)}{\overset{\psi_{-1/2}^{r}%
}{{\bar{a}}^{i}{\bar{\xi}}^{r}}}{,~}\underset{(0,0,6)}{\overset{\phi^{\left[
rs\right]  }}{{\bar{\xi}}^{r}{\bar{\xi}}^{s}}}{,~}\underset{(0,\frac{1}%
{2},\bar{4})}{\overset{\psi_{+1/2,a}}{{\bar{b}}^{I}{\bar{\xi}}^{r}{\bar{\xi}%
}^{s}{\bar{\xi}}^{m}}}{,~}\underset{(0,1,1)}{\overset{A_{+1}}{{\bar{b}}%
^{I}{\bar{b}}^{J}{\bar{\xi}}^{r}{\bar{\xi}}^{s}{\bar{\xi}}^{m}{\bar{\xi}}^{n}%
}}\right)  |0\rangle \label{sym}
\end{equation}
The notation in the last line is $\ \left(  j_{1},j_{2},\dim\left(
\text{SU}\left( 4\right)  \right)  \right)  ,\;$where $\left(
j_{1},j_{2}\right) ~$is for SU$\left(  2\right)  \times$SU$\left(
2\right)  $ while $\dim\left( \text{SU}\left(  4\right)  \right)  $
is the dimension of the SU$\left( 4\right)  $ representation. In
arriving at the SU$\left(  2\right)  \times $SU$\left(  2\right)
\times$SU$\left( 4\right)  $ representation labels in the third
line, we took into account that ${\bar{a}}^{i}{\bar{a}}^{j}$ is
symmetric while {$\bar{\xi}$}$^{r}${$\bar{\xi}$}$^{s}$ is
antisymmetric, etc. All other states with $\Delta=2$ are descendants
of these, and are obtained by applying arbitrary powers of the
double creation generator $\bar{a}^{j}\bar {b}^{I}$ in SU$\left(
2,2\right)  .$ All states have positive norm by virtue of the
positive norm oscillators we identified above. So, the towers of
states generated on each lowest state is an irreducible infinite
dimensional \textit{unitary} representation of SU$\left(  2,2\right)
$. The full collection of states is a single irreducible
representation of PS$\left( 2,2|4\right)  $ called the doubleton
representation of PSU$\left( 2,2|4\right)  $ (sometimes it is also
called the singleton, so the name is not so important).

We have shown that the list above is equivalent to a classification under
SU$\left(  2,2\right)  \times$ SU$\left(  4\right)  ,$ so the lowest states
should be sufficient to identify the SYM fields, and the descendants should be
analogous to applying multiple derivatives on a field since $\bar{a}^{j}%
\bar{b}^{I}$ is a vector (1/2,1/2) under SU$\left(  2\right)  \times
$SU$\left(  2\right)  .$ Indeed, we can imagine now an analytic continuation
back to the SL$\left(  2,C\right)  $ basis instead of the SU$\left(  2\right)
\times$SU$\left(  2\right)  $ basis and reinterpret the $\left(  j_{1}%
,j_{2}\right)  $ as the SL$\left(  2,C\right)  $ labels for the field. In this
analytic continuation the spin subgroup SO$\left(  3\right)  $ is a common
subgroup in both SL$\left(  2,C\right)  =$SO$\left(  3,1\right)  $ and
SU$\left(  2\right)  \times$SU$\left(  2\right)  =$SO$\left(  4\right)  ,$
therefore the spin of the state is $spin=j_{1}+j_{2}.$ The helicity and
chirality in SL$\left(  2,C\right)  $ are related, so by using the spin and
chirality we can identify the helicity. Using this, in Eq.(\ref{sym}) we have
identified the SYM fields with their helicities $h$ above each of the
oscillator combination. Although we have indicated the gauge field as
$A_{\pm1},$ as if it is only two states, the full SU$\left(  2\right)  \times
$SU$\left(  2\right)  $ or SL$\left(  2,C\right)  $ set of oscillator states
(1,0) and (0,1) really correspond to all the 6 components of the gauge
invariant field strength $F_{\mu\nu}$ in SL$\left(  2,C\right)  $ notation.
These comments are consistent with the helicities of the $A_{\mu}$ versus
$F_{\mu\nu}$ listed in Eq.(\ref{Zk}). Similar comments apply to all the other
spin $\left(  j_{1},j_{2}\right)  $ multiplets.

Although we gave a list of lowest states above as a supermultiplet, there
really is only one lowest oscillator state for the entire unitary
representation of PSU$\left(  2,2|4\right)  $. That one is simply {$\bar{\xi}%
$}$^{r}${$\bar{\xi}$}$^{s}|0\rangle,$ which satisfies $\Delta=2$ and is
annihilated not only by $a_{i}b_{J}$ but also by the supersymmetry generators
$\xi_{r}b_{J}$, $a_{i}\bar{\xi}^{s}$ that are part of $J.$ This is the lowest
state from which all other states with $\Delta=2$ listed can be obtained as
descendants by applying all powers of $J_{A}^{B}$ on this state (note
$[\Delta,J_{A}^{~B}]=0$). This entire tower is the doubleton of PSU$\left(
2,2|4\right)  $. If we watch carefully the orders of the oscillators we can
show that the generators $J=Z\bar{Z}$ of PSU$\left(  2,2|4\right)  $
\textit{in the doubleton representation} satisfy \cite{2tAdSs}%
\cite{2tsuperstring} the following nonlinear constraints as listed in Table 2.%
\begin{equation}
\left(  JJ\right)  _{A}^{~B}=-2\left(  J\right)  _{A}^{~B}+0\delta_{A}^{~B}.
\label{JJ}%
\end{equation}
The linear $J$ follows from the commutation rules among the generators, the
coefficient $-2$ is related to the commutation rules among the $J$'s but also
to the overall normalization of $J$ (taken differently in \cite{2tAdSs}%
\cite{2tsuperstring}), while the coefficient $0$ is the PSU$\left(
2,2|4\right)  $ quadratic Casimir eigenvalue $C_{2}=0$. We see that
the renormalized operator $\left(  -J/2\right)  $ acts as a
projection operator on physical states. This equation should be
viewed as a set of constraints on the generators that are satisfied
only in the doubleton representation, and as such this relation
identifies uniquely the representation only in terms of the
generators $J$. If the theory is expressed in any other form (such
as particle description, or field theory) the doubleton
representation can be identified in terms of the global symmetry as
one that must satisfy the constraints (\ref{JJ}), automatically
requiring the 6 scalars $\phi^{\left[  ab\right]  }$ as the lowest
SU$\left(  4\right)  $ multiplet. This is a completely PSU$\left(
2,2|4\right)  $ covariant and gauge invariant way of identifying the
physical and unitary states of the theory. Of course, the $d=4,$
$N=4$ SYM fields satisfy this criterion as seen above.

\subsection{Twistors for $d=4,$ $N=8$ SUGRA}

We can repeat the $N=4$ analysis of the superparticle or twistors for other
values of $N,$ and still $d=4.$ Twistor or 1T-superparticle are different
gauge choices of the 2T superparticle action $S_{2T}\left(  X,P,g\right)  ,$
so we expect the same physical spectrum. In the 1T-superparticle gauge of
Eq.(\ref{sparticle}) we start out with $4N$ real $\theta$'s in the action, but
due to kappa supersymmetry only $2N$ real $\theta$'$s$ can form physical
states in the lightcone gauge. The quantum algebra among the physical $\theta
$'s is the 2N dimensional Clifford algebra which is equivalent to N creation
and N annihilation fermionic oscillators. With the creation operators we
construct $2^{N}$ physical states, half are bosons and the other half are
fermions. Therefore for the superparticle the lightcone spectrum is
2$_{\text{bose }}^{N-1}$+2$_{\text{fermi}}^{N-1}.$

Since each fermionic $\theta$ carries spin $1/2,$ and SU$\left(  N\right)  $
quantum numbers in the fundamental representation, we can obtain the SU$(N)$
and helicity quantum numbers of the physical sates by assigning a Young
tableau $\square_{1/2}$ for the $N$ creation operators with helicity $1/2,$
and take antisymmetric products (fermions) to construct the 2$_{\text{bose }%
}^{N-1}$+2$_{\text{fermi}}^{N-1}$ physical states. If we start from a helicity
$-h$ for the $\theta$ vacuum, then the physical states have helicity and
SU$\left(  N\right)  $ quantum numbers given by the following Young tableaux
(this is for $N$=even$)$%
\[
\text{even:\ \ }\left(  0_{-h}+_{\square-h+1}^{\square}\cdots
+_{_{{\Large \square}}^{{\Large \square}}-h+\frac{N}{2}}^{_{{\Large \vdots}%
}^{{\Large \square}}}\right)  ;\;\text{odd:~}\left(  {\Large \square
}_{-h+\frac{1}{2}}+_{^{{\Large \square}}-h+\frac{3}{2}}^{_{{\Large \square}%
}^{{\Large \square}}}\cdots+_{_{{\Large \square}}^{{\Large \square}}%
}^{_{{\Large \vdots}}^{{\Large \square}}}{}_{-h+\frac{N-1}{2}}\right)
\]
A CPT invariant spectrum emerges provided for the even number of boxes
(similarly for the odd) the top helicity is the opposite of the bottom
helicity $-h+\frac{N}{2}=h.$ So the top and bottom helicities are $h=\pm N/4$.
This applies for $N$=even, and is consistent with half integer quantized
helicity. If $N$ is odd we do not get a CPT invariant spectrum. So, let us
consider the even $N=2,4,6,8$ cases. For $N=4,8$ the even number of boxes have
integer helicities hence they are bosons, and the odd number of boxes have
half-integer helicities hence they are fermions. For $N=2,6,$ the even number
of boxes are fermions, and the odd number of boxes are bosons.

Besides the $N=4$ case that gave the SYM spectrum, a most interesting case is
$N=8.$ This gives the top/bottom helicities for the bosons $h=\pm(8/4)=\pm2,$
which correspond to the graviton. Therefore the full spectrum gives the $N=8,$
$d=4$ supergravity spectrum.

We now analyze the theory in Eq.(\ref{Stw}), for $d=4,$ and any $N$ from the
point of view of twistors. For general $N$ the quantum ordered constraint is%
\begin{equation}
\frac{1}{2}\langle Z|\left(  Z_{A}\bar{Z}^{A}+\left(  -1\right)  ^{A}\bar
{Z}^{A}Z_{A}\right)  |\phi\rangle=0,
\end{equation}
where $\left(  -1\right)  ^{A}=\pm1$ is inserted for the bose/fermi components
since $Z_{A}$ is a PSU$\left(  2,2|N\right)  $ supertwistor. In $Z$ space this
is rearranged to the form
\begin{equation}
Z_{A}\frac{\partial\phi\left(  Z\right)  }{\partial Z_{A}}+\frac{1}%
{2}Str\left(  1\right)  \phi\left(  Z\right)  =0,\;Str\left(  1\right)  =4-N.
\end{equation}
We write $Z_{A}=\binom{z_{i}}{\xi_{a}}$ with $z_{i}=\binom{\mu}{\lambda}$ the
$4$ of SU$\left(  2,2\right)  \subset$PSU$\left(  2,2|N\right)  $ and $\xi
_{a}$ the $N$ of SU$\left(  N\right)  \subset$PSU$\left(  2,2|N\right)  .$
Then the superfield $\phi\left(  Z\right)  $ can be expanded in powers of the
fermions $\xi_{a}$%
\begin{equation}
\phi\left(  Z\right)  =\sum_{n=0}^{N}\left(  \xi_{a_{1}}\cdots\xi_{a_{n}%
}\right)  \phi^{a_{1}\cdots a_{n}}\left(  z\right)  .
\end{equation}
The equation $\frac{1}{2}\left(  N-4\right)  \phi\left(  Z\right)  =Z_{A}%
\frac{\partial\phi\left(  Z\right)  }{\partial Z_{A}}=z_{i}\frac{\partial
\phi\left(  z,\xi\right)  }{\partial z_{i}}+\xi_{a}\frac{\partial\phi\left(
z,\xi\right)  }{\partial\xi_{a}}=0$ becomes a homogeneity condition for the
coefficients $\phi^{a_{1}\cdots a_{n}}\left(  z\right)  $
\begin{equation}
z_{i}\frac{\partial\phi^{a_{1}\cdots a_{n}}\left(  z\right)  }{\partial z_{i}%
}=-\left(  n+\frac{4-N}{2}\right)  \phi^{a_{1}\cdots a_{n}}\left(  z\right)
\end{equation}
Comparing to Eq.(\ref{htwistor}) we see that the helicity of the wavefunction
$\phi^{a_{1}\cdots a_{n}}\left(  z\right)  $ is $h_{n}=\frac{n}{2}-\frac{N}%
{4}.$ So, for a given $N$ the lowest helicity is $h_{\min}=-\frac{N}{4}$ and
the top helicity is $h_{\min}=\frac{N}{4}.$ This is consistent with
quantization of spin only for $N$=even. For $N=8$ we obviously get the gravity
supermultiplet in the form of a field in twistor space
\begin{align}
\phi\left(  Z\right)   &  =g_{-2}\left(  z\right)  +\xi_{a}\lambda_{-3/2}%
^{a}\left(  z\right)  +\xi_{a}\xi_{b}V_{-1}^{ab}\left(  z\right)  +\frac
{\xi_{a}\xi_{b}\xi_{c}}{3!}\psi_{-1/2}^{abc}\left(  z\right) \\
&  +\frac{\xi_{a}\xi_{b}\xi_{c}\xi_{d}}{4!}~\phi_{0}^{abcd}\left(  z\right)
+\cdots+\xi^{8}g_{+2}\left(  z\right)  ,\nonumber\\
&  =1_{-2}+8_{-3/2}+28_{-1}+56_{-1/2}+70_{0}+56_{1/2}+28_{+1}+8_{+3/2}%
+1_{+2}\nonumber
\end{align}
where in the last line the SU$\left(  8\right)  $ representation and the
helicity are indicated. This is in agreement with the superparticle spectrum
discussed above.

If we are interested in a physical state $|\phi\rangle$ of definite momentum
$|k\rangle,$ we can use the table in Eq.(\ref{Zk}) to write the wavefunction
for each component of the superfield above, for $\phi_{h}\left(  z\right)
=\left(  g_{-2}\left(  z\right)  ,\cdots,g_{-2}\left(  z\right)  \right)  $
\begin{equation}
\phi_{h}\left(  z\right)  =\delta\left(  \langle\lambda\pi\rangle\right)
\exp\left( - \frac{\pi}{\lambda}\bar{\pi}_{\dot{\alpha}}\mu^{\dot{\alpha}%
}\right)  \left(  \frac{\lambda}{\pi}\right)  ^{-1-2h}\phi_{h}\left(  \pi
,\bar{\pi}\right)
\end{equation}

We can also approach the twistor quantum theory from the point of view of
oscillators$.$The formalism of Eqs.(\ref{oscillators}-\ref{oscillatorJ})
applies just by taking $N=8$ instead of $N=4.$ Then we obtain the spectrum of
lowest states in Fock space with $\Delta=4$ as follows%
\[
\left(  \underset{\left(  2,0,1\right)  }{\overset{g_{-2}}{\bar{a}^{4}}%
},\;\underset{\left(  \frac{3}{2},0,8\right)  }{\overset{\psi_{-3/2}^{a}}%
{\bar{a}^{3}\bar{\xi}}},\;\underset{\left(  1,0,28\right)  }{\overset
{A_{-1}^{\left[  ab\right]  }}{\bar{a}^{2}\bar{\xi}^{2}}},\;\underset{\left(
\frac{1}{2},0,56\right)  }{\overset{\lambda_{-1/2}^{\left[  abc\right]  }%
}{\bar{a}\bar{\xi}^{3}}},\;\underset{\left(  0,0,70\right)  }{\overset
{\phi^{\left[  abcd\right]  }}{\bar{\xi}^{4}}},\;\underset{\left(  0,\frac
{1}{2},56\right)  }{\overset{\lambda_{+1/2}^{\left[  abc\right]  }}{\bar
{b}\bar{\xi}^{5}}},\;\underset{\left(  0,1,28\right)  }{\overset
{A_{+1}^{\left[  ab\right]  }}{\bar{b}^{2}\bar{\xi}^{6}}},\;\underset{\left(
0,\frac{3}{2},8\right)  }{\overset{\psi_{+3/2}^{a}}{\bar{b}^{3}\bar{\xi}^{7}}%
},\;\underset{\left(  0,2,1\right)  }{\overset{g_{+2}}{\bar{b}^{4}\bar{\xi
}^{8}}}\right)
\]
As explained in the case of $N=4,$ the representations under $g_{\mp2}$
labeled as $\left(  2,0,1\right)  $ and $\left(  0,2,1\right)  $ are really
the components of the gauge invariant curvature tensor $R_{\alpha\beta
\gamma\delta}$ and $R_{\dot{\alpha}\dot{\beta}\dot{\gamma}\dot{\delta}}$ in
agreement with the table in Eq.(\ref{Zk}). Similar comments apply to the other
states labeled as $\left(  j_{1},j_{2},\dim\left(  \text{SU}\left(  8\right)
\right)  \right)  .$

\subsection{Supertwistors for d=6 and self dual supermultiplet}

The superparticle in $d=6$ and $N=4,$ derived from the 2T-physics theory
$S\left(  X,P,g\right)  $ as in Eq.(\ref{sparticle}), starts out with
$6x,6p,16\theta$ real degrees of freedom. Fixing $\tau,$ and kappa local
gauges and solving constraints, reduces the physical degrees of freedom down
to 5$x$, 5$p$, 8 $\theta.$ The superparticle action has a hidden global
superconformal symmetry OSp$(8^{\ast}|4)$ \cite{2tSuper}\cite{2ttwistor},
therefore the physical states should be classified as a unitary representation
under this group \cite{2tstringtwistors}.

If we quantize in the lightcone gauge we find $8_{B}+8_{F}$ states, which
should be compared to the physical fields of a six dimensional field theory
\begin{equation}
\underset{\text{self dual }F_{[\mu\nu\lambda]}^{+}=\partial_{\lbrack\mu_{1}%
}A_{\mu_{2}\mu_{3}]}=\varepsilon_{\mu_{1}\mu_{2}\mu_{3}\mu_{4}\mu_{5}\mu_{6}%
}\partial^{\lbrack\mu_{4}}A^{\mu_{5}\mu_{6}]}}{\text{{\ SO(5,1)}}%
\times\text{{Sp(4):}}\;F_{[\mu\nu\lambda]}^{+},\;\psi_{\alpha}^{a}%
,\;\phi^{\lbrack ab]}} \label{6Dfields}%
\end{equation}
taken in the lightcone gauge. Indeed we have the following 8 bosonic fields in
the lightcone gauge: a self dual antisymmetric tensor $A_{ij}=i\varepsilon
_{ijkl}A^{kl}$ in SO$\left(  4\right)  \subset$SO$(5,1)$ (i.e. 3 fields), and
the Sp$\left(  4\right)  $ traceless antisymmetric $\phi^{\lbrack ab]}$ (5
scalars). Half of the $4\times4$ pseudoreal $\psi_{\alpha}^{a}$ are
independent degrees of freedom on the lightcone, and can be classified under
SU$\left(  2\right)  \times$SU$\left(  2\right)  \times$ Sp$\left(  4\right)
$ as 8 pseudo-real fermionic fields $\left(  2,0,4\right)  , $ consistent with
expectation. An \textit{interacting} quantum conformal field theory with these
degrees of freedom is expected but cannot be written down covariantly in the
form of a local field theory. The twistor approach may be helpful.%

\[%
\begin{tabular}
[c]{|l|}\hline
$Z_{Aa}=\left(
\genfrac{}{}{0pt}{}{{8bose}}{{4fermi}}%
\right)  \;\;\;%
\genfrac{}{}{0pt}{}{\text{{12x2 rectangular matrix,\ }}{A=1,\cdots
,12;\;}\text{{\ a}}=1,2}{\text{{2 twistors in fundamental rep of }}%
{OSp(8}^{\ast}{|4)}}%
$\\\hline
$Z_{Aa}~=\text{{\ (12,2)} {\ of }}{OSp(8}^{\ast}{|4)}_{\text{global}}%
\times\text{{\ SU(2)}}_{\text{local}}$\\\hline
$L=\bar{Z}^{Aa}\left(  \left(  \partial+V\right)  Z\right)  _{Aa}%
,\;V=\text{SU}\left(  2\right)  \text{ gauge field}$\\\hline
$%
\genfrac{}{}{0pt}{}{\text{{Pseudo-real~~~~~~~~}}}{\text{1st \&~2nd~related}}%
\text{\ }Z_{Aa}=\left(
\begin{array}
[c]{cc}%
a_{{1}i} & a_{{2}i}\\
\bar{a}_{{2}}^{i} & -\bar{a}_{{1}}^{i}\\
\xi_{{1}\alpha} & \xi_{{2}\alpha}\\
\bar{\xi}_{{2}}^{\alpha} & -\bar{\xi}_{{1}}^{\alpha}%
\end{array}
\right)  \;%
\genfrac{}{}{0pt}{}{i:~~\text{4~of}~~\text{SU}(4)=\text{SO}\left(  6\right)
\subset\text{SO}(6,2)}{\alpha:\;\text{2~of}~~\text{SU}(2)\subset
\text{Sp}(4)\;\;~~~\;\;\;\;\;\;}%
$\\\hline
$\bar{Z}^{aA}=\left(  Z^{\dagger}\eta\right)  ^{aA}=\left(
\begin{array}
[c]{cccc}%
\bar{a}_{{1}}^{i} & -a_{{2i}} & \bar{\xi}_{{1}}^{\alpha} & \xi_{{2}\alpha}\\
\bar{a}_{{2}}^{i} & a_{{1i}} & \bar{\xi}_{{2}}^{\alpha} & -\xi_{{1}\alpha}%
\end{array}
\right)  ,\;\eta=diag\left(  1_{4},-1_{4},1_{2},1_{2}\right)  $\\\hline
$Z_{Aa}$ is pseudo real, $\bar{Z}^{aA}=\varepsilon^{ab}\left(  Z^{T}\right)
_{bB}C^{BA},\;C=\left(
\begin{array}
[c]{cc}%
\sigma_{1}\times1_{4} & 0\\
0 & -i\sigma_{2}\times1_{2}%
\end{array}
\right)  $\\\hline
\end{tabular}
\ \ \ \
\]

Let us now examine the twistors that emerge in Eq. (\ref{Stw}) for this case.
Writing the twistor in the oscillator basis we have the results listed above.
The pseudo-reality property follows from the fact that $Z_{Aa}$ is part of the
group element $g\in$OSp$\left(  8^{\ast}|4\right)  $ that satisfies
$g^{-1}=C^{-1}g^{ST}C$ with the $C^{BA}$ used above. Then $Z_{Aa}$ takes the
form above in a natural basis. Thus the second column is related to the first
one, but still consistent with a local SU$\left(  2\right)  $ applied on
$a=1,2.$

When $Z,\bar{Z}$ of these forms are inserted in the Lagrangian $L=\bar{Z}%
^{Aa}\left(  \left(  \partial+V\right)  Z\right)  _{Aa}$ we get (after
integration by parts and dropping total derivatives, and taking care of
bose/fermi statistics in interchanging factors)%
\begin{align*}
L  &  =\bar{a}_{{1}}^{i}\partial a_{{1}i}+\bar{a}_{{2}}^{i}\partial a_{{2}%
i}+\bar{\xi}_{{1}}^{\alpha}\partial\xi_{{1}\alpha}+\bar{\xi}_{{2}}^{\alpha
}\partial\xi_{{2}\alpha}-\frac{1}{2}Tr\left(  VG\right)  ,\;V=\text{SU}\left(
2\right)  \text{ gauge field.}\\
G  &  \equiv\left(  \bar{Z}Z\right)  _{a}^{~b}=\text{2}\times2\text{ traceless
SU}\left(  2\right)  \text{ gauge generator. }G\text{ = 0 on physical
states,}\\
&  =\left(
\begin{array}
[c]{cc}%
\bar{a}_{{1}}^{i}a_{{1}i}-a_{{2}i}\bar{a}_{{2}}^{i}+\bar{\xi}_{{1}}^{\alpha
}\xi_{{1}\alpha}+\xi_{{2}\alpha}\bar{\xi}_{{2}}^{\alpha}~~ & \bar{a}_{{1}}%
^{i}a_{{2}i}+a_{{2}i}\bar{a}_{{1}}^{i}+\bar{\xi}_{{1}}^{\alpha}\xi_{{2}\alpha
}-\xi_{{2}\alpha}\bar{\xi}_{{1}}^{\alpha}\\
\bar{a}_{{2}}^{i}a_{{1}i}+a_{{1}i}\bar{a}_{{2}}^{i}+\bar{\xi}_{{2}}^{\alpha
}\xi_{{1}\alpha}-\xi_{{1}\alpha}\bar{\xi}_{{2}}^{\alpha}~~ & \bar{a}_{{2}}%
^{i}a_{{2}i}-a_{{1}i}\bar{a}_{{1}}^{i}+\bar{\xi}_{{2}}^{\alpha}\xi_{{2}\alpha
}+\xi_{{1}\alpha}\bar{\xi}_{{1}}^{\alpha}%
\end{array}
\right) \\
J  &  =Z\bar{Z}-\frac{1}{4}Str\left(  Z\bar{Z}\right)  ,\text{ 12}%
\times12\text{ supermatrix of global OSp}\left(  8^{\ast}|4\right)  \text{
charges.}%
\end{align*}
It is seen that according to the canonical formalism, the oscillators
identified above all have positive norm%
\begin{equation}
\left[  a_{1i},\bar{a}_{1}^{j}\right]  =\delta_{i}^{j}=\left[  a_{2i},\bar
{a}_{2}^{j}\right]  ,\;\;\left\{  \xi_{{1}\alpha},\bar{\xi}_{{1}}^{\beta
}\right\}  =\delta_{\alpha}^{\beta}=\left\{  \xi_{{2}\alpha},\bar{\xi}_{{2}%
}^{\beta}\right\}  .
\end{equation}
We count the degrees of freedom before the constraints, and find that $Z_{Aa}$
has $(8_{B}+4_{F})\times2=16_{B}+8_{F}$ real parameters (namely the complex
$a_{{1}i},a_{{2}i},\xi_{{1}\alpha},\xi_{{2}\alpha}$). The constraints are due
to a SU$\left(  2\right)  $ gauge symmetry acting on the index $a=1,2$
(although it seems like SU$\left(  2\right)  \times$U$\left(  1\right)  ,$ the
U$\left(  1\right)  $ part is automatically satisfied because of the
pseudoreal form of $Z_{Aa}$). The 3 gauge parameters and 3 constraints remove
6 bosonic degrees of freedom, and we remain with 10$_{B}+8_{F}$ physical
degrees of freedom. This is the same as the count for the superparticle
($5x,5p,8\theta$). It is obvious we have the same number of degrees of freedom
and the same symmetries OSp$\left(  8^{\ast}|4\right)  ,$ with the symmetry
being much more transparent in the twistor basis.

The quantum theory can proceed in terms of the oscillators. There is only one
lowest physical state, namely the Fock vacuum $|0\rangle$ that is gauge
invariant $G|0\rangle=0,$ and is also annihilated by the double annihilation
operators of SO$\left(  6,2\right)  $ in $J=Z\bar{Z}$. All other $G=0$
physical states are descendants of this one by applying all powers of $J.$ The
resulting representation is precisely the doubleton of OSp$\left(  8^{\ast
}|4\right)  $ which is equivalent to the fields in Eq. (\ref{6Dfields}) and
Table 3 below. This oscillator representation was worked out long ago in
\cite{gunaydinWarner} using again the Bars-G\"{u}naydin method
\cite{barsgunaydin}.

The details of the doubleton are found as follows. First construct the
SU$\left(  2\right)  $ gauge singlet (i.e.. $G=0$) ground states in Fock space
as in the first column of Table 3. The ground states are the states
annihilated by the double annihilation bosonic generator $\left(  aa\right)
=a_{ai}a_{bj}\varepsilon^{ab}=a_{1i}a_{2j}-a_{1j}a_{2i}.$ This is one of the
generators in $J=Z\bar{Z}-\frac{1}{4}Str\left(  Z\bar{Z}\right)  $ that sits
in the conformal subgroup SO$^{\ast}\left(  8\right)  $ (i.e. spinor of
SO$\left(  6,2\right)  $). All the states that are annihilated by this
generator are included in the first column below (note zero or one power of
$\bar{a}$ is obviously annihilated, the two powers of $\bar{a}$ in the last
item is possible only because of an appropriate symmetrization as described
below, higher powers of $\bar{a}$ cannot be annihilated by this generator if
we also require annihilation by $G$)
\begin{align}
&
\begin{tabular}
[c]{|l|l|l|l|}\hline
$%
\genfrac{}{}{0pt}{}{\text{Fock space}}{\text{lowest state}}%
$ & $%
\genfrac{}{}{0pt}{}{\text{SU(4)}\times\text{SU(2)}}{\text{Young tableau}}%
$ & $%
\genfrac{}{}{0pt}{}{\text{SU(4)}\times\text{Sp(4)}}{\text{dimensions}}%
$ & $%
\genfrac{}{}{0pt}{}{\text{SO(5,1)}\times\text{Sp(4)}}{\text{field}}%
$\\\hline
$%
\begin{tabular}
[c]{l}%
$|0\rangle$\\
$\left(  \bar{\xi}\bar{\xi}\right)  |0\rangle$\\
$\left(  \bar{\xi}\bar{\xi}\right)  ^{2}|0\rangle$%
\end{tabular}
\ \ $ & $%
\begin{tabular}
[c]{l}%
$\left(  0,0\right)  \;\;=\left(  1,1\right)  $\\
$\left(  0,{\scriptsize \square\square}\right)  =\left(  1,3\right)  $\\
$\left(  0,_{\square\square}^{\square\square}\right)  =\left(  1,1\right)  $%
\end{tabular}
\ \ $ & $\left(  1,5\right)  $ & $\phi^{\left[  ab\right]  }$\\\hline
$%
\begin{tabular}
[c]{l}%
$\left(  \bar{a}\bar{\xi}\right)  |0\rangle$\\
$\left(  \bar{\xi}\bar{\xi}\right)  \left(  \bar{a}\bar{\xi}\right)
|0\rangle$%
\end{tabular}
\ \ $ & $%
\begin{tabular}
[c]{l}%
$\left(  {\small \square,\square}\right)  \;=\left(  4,2\right)  $\\
$\left(  {\small \square},_{\square}^{\square\square}\right)  =\left(
4,2\right)  $%
\end{tabular}
\ \ $ & $\left(  4,4\right)  $ & $\psi_{\alpha}^{a}$\\\hline
$\left(  \bar{a}\bar{\xi}\right)  \left(  \bar{a}\bar{\xi}\right)  |0\rangle$
& $\left(  {\scriptsize \square\square},_{\square}^{\square}\right)  =\left(
10,1\right)  $ & $\left(  10,1\right)  $ & $\partial_{\lbrack\lambda}A_{\mu
\nu]^{+}}~%
\genfrac{}{}{0pt}{}{\text{self}}{\text{dual}}%
$\\\hline
\end{tabular}
\nonumber\\
&  \text{Table 3 - The OSp}\left(  8^{\ast}|4\right)  \text{ doubleton.}%
\end{align}
To insure that these are also annihilated by the SU$\left(  2\right)  $ gauge
generators $G_{a}^{~b},$ we must combine the SU$\left(  2\right)  $ gauge
indices on the oscillators into SU$\left(  2\right)  $ singlets. This
guarantees that all of these states are physical. So $\left(  \bar{\xi}%
\bar{\xi}\right)  ,\left(  \bar{a}\bar{\xi}\right)  $ and the creation
generator $\left(  \bar{a}\bar{a}\right)  $ stand for
\begin{equation}
\left(  \bar{\xi}\bar{\xi}\right)  \equiv\bar{\xi}_{a}^{\alpha}\bar{\xi}%
_{b}^{\beta}\varepsilon^{ab}=\left(  0,{\scriptsize \square\square}\right)
,\;\left(  \bar{a}\bar{\xi}\right)  \equiv\bar{a}_{a}^{i}\bar{\xi}_{b}^{\beta
}\varepsilon^{ab}=\left(  {\small \square,\square}\right)  ,\;\left(  \bar
{a}\bar{a}\right)  =\bar{a}_{a}^{i}\bar{a}_{b}^{j}\varepsilon^{ab}=\left(
_{\square}^{\square},0\right)  \label{yt}%
\end{equation}
while the annihilation $\left(  aa\right)  $ generator stands for $\left(
aa\right)  =a_{ai}a_{bj}\varepsilon^{ab}=\left(  _{\boxdot}^{\boxdot
},0\right)  ,$ where a dotted box represents the complex conjugate
representation (but for SU$\left(  4\right)  ,$ $_{\boxdot}^{\boxdot
}=_{\square}^{\square}).$ The boxes in the Young tableaux represent the
un-summed indices $i,\alpha$ which stand for the fundamental representations
of SU(4)$\times$SU(2) where SU(4)$\subset$SO$^{\ast}\left(  8\right)  $ and
SU(2)$\subset$Sp$\left(  4\right)  .$ To keep track of these indices we use
the Young tableaux notation as in Eq.(\ref{yt}) for the operators, and use
that property to figure out the second column of Table 3. In both
Eq.(\ref{yt}) and Table 3 we take into account that the $a$ oscillators are
bosons and the $\xi$ oscillators are fermions, so under permutations of the
$a$'s there must be symmetry and under permutation of the $\xi$'s there must
be antisymmetry. These properties lead uniquely to the Young tableaux listed
in the table. Next, for each SU$\left(  4\right)  $ representation we combine
the SU$\left(  2\right)  $ representations into complete Sp$\left(  4\right)
$ representations as in the third column of Table 3. From this we can easily
read off the corresponding fields as in the last column of Table 3. Having
established all possible ground states for the operator $\left(  aa\right)  ,$
we apply all possible powers of the generator $\left(  \bar{a}\bar{a}\right)
$ on those ground states in order to obtain all the states of the irreducible
representation. Applying the powers of $\left(  \bar{a}\bar{a}\right)  $ just
gives the descendants of the ground states. The collection of all these states
is the same as starting with the single ground state $|0\rangle=\left(
0,0\right)  $ and then applying all the powers of the OSp$\left(  8|4\right)
$ generators $J=Z\bar{Z}-\frac{1}{4}Str\left(  Z\bar{Z}\right)  .$ The reason
for organizing the states as in Table 3, is to read off the standard
supersymmetry multiplet that corresponds to the fields in field theory as
described in the next paragraph (Poincar\'{e} supersymmetry is a subgroup of
OSp$\left(  8|4\right)  $).

In the last step of Table 3 we interpret SU$\left(  4\right)  $ as the
spacetime SO$\left(  5,1\right)  $ after an analytic continuation. The number
of states in the field theory notation must match the number of states in the
third column. For example the 5 states $\left(  1,5\right)  $ corresponds to
$\phi^{\left[  ab\right]  }$ which is a 4$\times4$ antisymmetric, and
\textit{traceless} tensor under the Sp$\left(  4\right)  $ metric, which has
just 5 components. Also $\partial_{\lbrack\lambda}A_{\mu\nu]^{+}}$ is a
10-component, 3-index antisymmetric and self-dual tensor, using SO$\left(
5,1\right)  $ vector indices $\mu,\nu,\lambda,$ instead of the spinor indices
${\scriptsize \square\square}=10$. These fields form the basis for the
self-dual tensor supermultiplet under supersymmetry in 6 dimensions.

Of course they are also the basis of the infinite dimensional unitary
representation of OSp$\left(  8|4\right)  .$ The generators of the latter are
of course the $J=Z\bar{Z}-\frac{1}{4}Str\left(  Z\bar{Z}\right)  ,$
constructed from the twistors in the form of a12$\times12$ supermatrix of
global OSp$\left(  8^{\ast}|4\right)  $ charges.

\section{2T superstring descends to twistor superstring}

So far in these lectures I discussed superparticles and the associated
supertwistors, and their physical spectra. These have a direct generalization
to superstrings via the 2T superstring formalism given in \cite{2tsuperstring}%
. Briefly, the action is $S=\int d\tau d\sigma\left(  L_{2T}^{+}+L_{2T}%
^{-}\right)  ,$ and $L_{2T}^{\pm}$ are defined on the worldsheet as follows%

\begin{align*}
L_{2T}^{\pm}  &  ={\partial}_{\pm}{X\cdot P}^{\pm}{-}\frac{1}{2}{AX\cdot
X-}\frac{1}{2}{B}_{\pm\pm}{P}^{\pm}{\cdot P}^{\pm}{-C}_{\pm}{P}^{\pm}{\cdot
X}\\
&  -\frac{1}{2^{[d/2]-1}}Str\left(  \partial_{\pm}g\bar{g}\left(
\genfrac{}{}{0pt}{}{L_{MN}^{\pm}\Gamma^{MN}}{0}%
\genfrac{}{}{0pt}{}{0}{0}%
\right)  \right)  +\mathit{L}_{G}%
\end{align*}
$X_{M}(\tau,\sigma),P_{M}^{\pm}(\tau,\sigma),$\ $L_{MN}^{\pm}=X_{[M}%
P_{M]}^{\pm},~g(\tau,\sigma)$ are now string fields, and
${\partial}_{\pm }=\frac{1}{2}\left(
\partial_{\sigma}\pm\partial_{\tau}\right)  .$ Here $L_{2T}^{\pm}$
represent left/right movers, and there is open string boundary
conditions$.$ $L_{G}$ describes additional degrees of freedom that
may be needed to insure a critical worldsheet theory that is
conformally exact. In the case of the $d=4$, $N=4$ twistor
superstring $L_{G}$ describes an internal current algebra for some
SYM group $G,$ with conformal central charge $c=28$ \cite{berko}.

The local and global symmetries are similar to those of the particle and are
described in \cite{2tsuperstring}. The global symmetry is $G_{\text{super}}$
chosen for various $d$ as in Table 1. One must insure that there are no
anomalies in the operator products of the local symmetry currents. Such
details will be discussed elsewhere.

In the twistor gauge the 2T superstring above reduces to twistor superstrings,
with the twistors described in the previous sections. In $d=4$, and
$G_{\text{super}}=$SU$\left(  2,2|4\right)  $ the 2T superstring descends to
the twistor superstring in the Berkovits version. There are open problems that
remain to be resolved in this theory \cite{berkWit}, in particular the
conformal supergravity sector is undesirable and some constraint that projects
out this sector would be interesting to find.

For other values of $d$ and $G_{\text{super}}$ these theories remain to be
investigated. We know of course the particle limit of these worldsheet
theories as discussed in these lectures. In particular the $d=4,$ $N=8$
twistors lead to $N=8$ supergravity in the particle limit, so this provides a
starting point for an investigation similar to \cite{witten2}\cite{cachazo} of
a twistor superstring for $d=4,$ $N=8$ supergravity.

Similarly to the $d+d^{\prime}=10,11$ particle case we also consider the
$d+d^{\prime}=10,11$ 2T superstrings%

\begin{align*}
\hat{L}_{2T}^{\pm}  &  ={\partial}_{\pm}{\hat{X}\cdot\hat{P}}^{\pm}{-}\frac
{1}{2}{A\hat{X}\cdot\hat{X}-}\frac{1}{2}{B}_{\pm\pm}{\hat{P}}^{\pm}{\cdot
\hat{P}}^{\pm}{-C}_{\pm}{\hat{P}}^{\pm}{\cdot\hat{X}}\\
&  -\frac{1}{8}Str\left(  \partial_{\pm}g\bar{g}\left(
\genfrac{}{}{0pt}{}{L_{MN}^{\pm}\Gamma^{MN}}{0}%
\genfrac{}{}{0pt}{}{0}{-\alpha L_{IJ}^{\pm}\Gamma^{IJ}}%
\right)  \right)
\end{align*}
where SO($\left(  d+d^{\prime}\right)  $,2)$\rightarrow$SO(d,2)$\times
$SO($d^{\prime}$), $\hat{X}^{M}=\left(  {X}^{m},{X}^{I}\right)  $, $\hat
{P}_{M}^{\pm}=\left(  {P}_{m}^{\pm},{P}_{I}^{\pm}\right)  $, $g(\tau,\sigma)$
$\in$ SU$\left(  {2,2|4}\right)  $ or OSp$\left(  8|4\right)  $, and
$L_{MN}^{\pm}=X_{[M}P_{M]}^{\pm}$, $L_{IJ}^{\pm}=X_{[I}P_{J]}^{\pm}$. The
local and global symmetries are discussed in the second part of section
(\ref{stwgaugeSec}). In the particle-type gauge, the spectrum in the particle
limit is the same as linearized type IIB SUGRA compactified on AdS$_{5}\times
$S$^{5}$ \cite{2tAdSs} for $d+d^{\prime}=10,$ and 11D SUGRA compactified to
AdS$_{4}\times$S$^{7}$ or AdS$_{7}\times$S$^{4}$ for $d+d^{\prime}=11$. In the
twistor gauge this theory is currently being investigated by using the
twistors in Eqs.(\ref{Zadss10},\ref{Zadss11}).

In the 2T philosophy each one of these 2T superstrings have many duals that
can be found and investigated by choosing various gauges. This is a completely
open field of investigation at this time, and it could be quite interesting
from the point of view of M-theory.

The analogies to certain aspects of M-theory are striking. Dualities in
M-theory appear to be analogs of the Sp(2,R) and its generalizations discussed
above, and illustrated in Fig.1 for the simplest model of 2T-physics. Taking
into consideration that 2T-physics correctly describes 1T-physics, and
provides a framework for a deeper view of spacetime and a new unification of
1T-dynamics, we are tempted to take the point of view that it probably applies
also to M-theory. So possibly M-theory would eventually be most clearly
formulated as a 13-dimensional theory with signature $(11,2)$ and global
supersymmetry OSp$(1|64)$. This is consistent with certain attractive features
of the supergroup OSp$(1|64)$ as a hidden symmetry of M-theory \cite{Stheory}%
-\cite{2ttoyM}.

The twistor approach discussed in these lectures may be a useful tool for
further progress in this deeper direction as well as for developing new
computational tools in conventional theories.\bigskip

\section{Acknowledgments}

I would like to thank the organizers of the \textquotedblleft2005 Summer
School on String/M Theroy\textquotedblright\ in Shanghai, China, and the
organizers of the International Symposium QTS4, \textquotedblleft Quantum
Theory and Symmetries IV\textquotedblright, in Varna, Bulgaria, for their
hospitality and support. This research was supported by the US Department of
Energy under grant No. DE-FG03-84ER40168.

\end{document}